\documentclass[modern, letterpaper]{aastex61}

\definecolor{cbblue}{HTML}{0D47A1}
\usepackage{microtype}  
\usepackage{amsmath,amssymb,natbib}
\usepackage[flushleft]{threeparttable}
\hypersetup{backref,breaklinks,colorlinks,urlcolor=cbblue,linkcolor=cbblue,citecolor=black}
\graphicspath{{figures/}}
\newcommand{\project}[1]{\textsl{#1}}
\newcommand{\acronym}[1]{{\small{#1}}}
\newcommand{\gaia}{\project{Gaia}}

\newcommand{\documentname}{Article}
\newcommand{\sectionname}{Section}
\newcommand{\figname}{Figure}
\newcommand{\eqname}{Equation}
\newcommand{\dr}{\acronym{DR1}}
\newcommand{\tgas}{\acronym{TGAS}}
\newcommand{\etal}{\textit{et al}.}
\newcommand*\elem[1]{\ensuremath{\mathrm{#1}}}
\newcommand*\elemH[1]{\ensuremath{[\mathrm{#1}/\elem{H}]}}
\newcommand*\teff{\ensuremath{T_\mathrm{eff}}}

\newcommand*{\feh}{\ensuremath{\elemH{Fe}}}
\newcommand{\sunanalog}{\text{Krios}}
\newcommand{\bizarreone}{\text{Kronos}}
\newcommand{\Tcondens}{\ensuremath{T_C}}
\newcommand{\mearth}{\ensuremath{M_\oplus}}
\newcommand{\mjupiter}{\ensuremath{M_\mathrm{Jup}}}

\newcommand{\transp}[1]{{#1}^{\!\mathsf{T}}}

\newcommand{\bs}[1]{\boldsymbol{#1}}

\newcommand{\mat}[1]{\mathbf{#1}}
\renewcommand{\vec}[1]{\bs{#1}}
\newcommand{\kms}{\ensuremath{\rm km~s^{-1}}}
\newcommand{\msun}{\ensuremath{{\mathrm M}_\odot}}
\newcommand{\pc}{{\rm pc}}

\newcommand{\maccreted}{\ensuremath{15~\mearth}}

\renewcommand\tablename{Table}

\begin{document}\sloppy\sloppypar\raggedbottom\frenchspacing 

\title{
  Kronos \& Krios:
  Evidence for accretion of a massive, rocky planetary system
  in a comoving pair of solar-type stars
}

\author[0000-0001-7790-5308]{Semyeong Oh}
\affil{Department of Astrophysical Sciences, Princeton University, 4 Ivy Lane,
  Princeton, NJ 08544, USA}
\affil{To whom correspondence should be addressed: \texttt{semyeong@astro.princeton.edu}}

\author[0000-0003-0872-7098]{Adrian M. Price-Whelan}
\affil{Department of Astrophysical Sciences, Princeton University, 4 Ivy Lane,
  Princeton, NJ 08544, USA}

\author[0000-0002-9873-1471]{John M. Brewer}
\affil{Department of Astronomy, Yale University, 260 Whitney Ave,
  New Haven, CT 06511, USA}
\affil{Department of Astronomy, Columbia University, 550 West 120th Street,
  New York, NY 10027, USA}

\author[0000-0003-2866-9403]{David W. Hogg}
\affil{Center for Computational Astrophysics, Flatiron Institute, 162 Fifth Ave,
  New York, NY 10010, USA}
\affil{Center for Cosmology and Particle Physics,
  Department of Physics, New York University, 726 Broadway,
  New York, NY 10003, USA}
\affil{Center for Data Science, New York University, 60 Fifth Ave,
  New York, NY 10011, USA}
\affil{Max-Planck-Institut f\"ur Astronomie, K\"onigstuhl 17, D-69117 Heidelberg}

\author[0000-0002-5151-0006]{David N. Spergel}
\affil{Department of Astrophysical Sciences, Princeton University, 4 Ivy Lane,
  Princeton, NJ 08544, USA}
\affil{Center for Computational Astrophysics, Flatiron Institute, 162 Fifth Ave,
  New York, NY 10010, USA}
\author{Justin Myles}
\affil{Department of Astronomy, Yale University, 260 Whitney Ave, New Haven, CT 06511, USA}

\begin{abstract}
  We report and discuss the discovery of a comoving pair of bright solar-type
  stars, HD~240430 and HD~240429, with a significant difference in their
  chemical abundances.
  The two stars have an estimated 3D separation of $\approx 0.6$~pc
  ($\approx 0.01$~pc projected) at a distance of
  $r\approx 100~\pc$ with nearly identical three-dimensional velocities,
  as inferred from \gaia\ \tgas\ parallaxes and proper motions, and
  high-precision radial velocity measurements.
  Stellar parameters determined from high-resolution Keck HIRES spectra
  indicate that both stars are $\sim 4$~Gyr old.
  The more metal-rich of the two, HD~240430, shows an enhancement of refractory
  ($\Tcondens>1200$~K) elements by $\approx 0.2$~dex and a marginal
  enhancement of (moderately) volatile elements ($\Tcondens<1200$~K; \elem{C},
  \elem{N}, \elem{O}, \elem{Na}, and \elem{Mn}).
  This is the largest metallicity difference found in a wide binary pair yet.
  Additionally, HD~240430 shows an anomalously high surface lithium abundance
  ($A(\elem{Li})=2.75$), higher than its companion by $0.5$~dex.
  The proximity in phase-space and ages between the two stars suggests that
  they formed together with the same composition, at odds with the observed
  differences in metallicity and abundance patterns.
  We therefore suggest that the star HD~240430, ``Kronos'', accreted
  \maccreted\ of rocky material after birth, selectively enhancing the
  refractory elements as well as lithium in its surface and convective
  envelope.
\end{abstract}

\keywords{
  binaries: visual
  ---
  planet-star interactions
  ---
  stars: abundances
  ---
  stars: formation
  ---
  stars: individual (HD~240430, HD~240429)
  ---
  stars: solar-type
}

\section{Introduction} 
\label{sec:introduction}

Wide binary stars are valuable tools for studying star and planet formation as
well as Galactic dynamics and chemical evolution.
In the context of studying the evolution of the Milky Way, they are useful for
two main reasons.
First, because wide binaries are weakly bound systems that may be tidally
disrupted by, e.g., field stars, molecular clouds, or the Galactic tidal field,
their statistics can be informative of the Galactic mass distribution.
For example, the separation distribution of halo binaries has been used to
constrain the mass of massive compact halo objects
(\citealt{Yoo:2004aa,Quinn:2009,Allen:2014}).
They can also be used to test the ``chemical tagging'' hypothesis that stars
from the same birthplace may be traced back using detailed chemical abundance
patterns as birth tags (\citealt{2002ARA&A..40..487F}).
While any multiple-star system, including massive open clusters, can be used to
test the hypothesis, wide binaries have the advantage of being extremely
abundant, rendering their statistics a meaningful indication of whether the
hypothesis works.

Binary stars that form from the same birth cloud start with nearly identical
composition.
A differential analysis of the chemical composition of binary stars can reveal
their history through the chemical signatures related to planet formation or
accretion regardless of Galactic chemical evolution.
Giant planets on short period orbits have been shown, via
population studies, to form more readily around inherently metal rich stars
(e.g., \citealt{Fischer:2005aa,Santos2004}).
However, the post-formation accretion of rocky planets can still alter the
photospheric abundances.
If host stars are polluted after their birth by rocky planetary material with a
high refractory-to-volatile ratio, the convective envelope of the stars may be
enhanced in refractory elements (e.g., \elem{Fe}) compared to their initial
state (e.g., \citealt{Pinsonneault:2001aa}).
Thus, differences in planet formation or accretion in two otherwise identical
stars may imprint differences in chemical abundances that depend on the
condensation temperature (\Tcondens).

High resolution spectroscopic studies of binary star systems hosting at least
one planet (\citealt{2011ApJ...740...76R,2014ApJ...790L..25T,Teske:2013aa,
  Mack:2014aa,Liu:2014aa,Teske:2015aa,Saffe:2015aa,
  Ramirez:2015aa,Biazzo:2015aa,Mack:2016aa,Teske:2016aa,Teske:2016ab})
have yielded varied results:
while some systems appear to have undetectable differences
(see also \citealt{Desidera:2004aa,Gratton:2001aa}), other
studies have reported a \Tcondens-dependent difference in abundance
with higher-\Tcondens\ elements showing larger differences.
A possible explanation for the difference is that forming more gas giants or
rocky planets leads to an overall or \Tcondens-dependent depletion of metals in
gas that eventually accretes onto the host star
(\citealt{Ramirez:2015aa,Biazzo:2015aa}).
Alternatively, late time accretion of refractory-rich planetary material can
also produce the trend by enhancing the abundance of high-\Tcondens\ elements
in one of the two stars.
The observed differences are $\lesssim 0.1$~dex even in the most dramatic case,
and often at a level of $\approx 0.05$~dex, making them challenging to detect even
with a careful analysis of high-resolution, high signal-to-noise ratio spectra,
and differential analyses of two stars that are very similar in their
stellar parameters.
We refer the readers to Appendix~\ref{app:review} for a review of
a handful of individual pairs studied in their detailed chemical abundances
(see also \citealt{2016arXiv161104064M}).

Spectral analysis of polluted white dwarfs (WDs) currently provides the
strongest evidence for accretion of planetary material by a host star
(\citealt{2003ApJ...596..477Z,2010ApJ...722..725Z,2014A&A...566A..34K};
see \citealt{2016NewAR..71....9F} for review).
Because the gravitational settling times of elements heavier than \elem{He} in
the WD atmosphere is much shorter than the WD cooling time
(\citealt{1986ApJS...61..197P}), detection of metals likely indicates the
presence of a reservoir of dusty material around the WD.
Indeed, many of the polluted WDs host a dusty debris disk detected in the
infrared (\citealt{1987Natur.330..138Z,1990ApJ...357..216G,2005ApJ...635L.161R,
  2009ApJ...694..805F,2006ApJ...646..474K}).
Some of the most dramatically polluted WDs show
surface abundances closely matched by rocky planetary material
with, e.g., bulk Earth composition, strongly arguing
that the disk formed from tidally disrupted minor planets
(\citealt{Zuckerman:2007aa,Klein:2010aa}).
Recently, transit signals from small bodies orbiting around a polluted WD
have been detected by \project{Kepler} adding further support to the picture
(\citealt{2015Natur.526..546V}).

Here, we report and discuss the discovery of a comoving pair of G stars,
HD~240430 and HD~240429, with unusual chemical abundance differences that
strongly suggest accretion of rocky planetary material by one of the two stars,
HD~240430.
Throughout the \documentname, we nickname the two stars \bizarreone\
(HD~240430) and \sunanalog\ (HD~240429).
In Greek mythology, Kronos and Krios were sons of Uranos and Gaia.
Kronos notoriously devoured all of his children (except Zeus)
to prevent the prophecy that one day he will be overthrown by them.
We use the following convention for chemical abundances of stars:
\elemH{X} is the log ratio of the number density of an element \elem{X} to \elem{H}
relative to the solar value,
$\elemH{X} = \log_{10} ((n_\elem{X}/n_\elem{H})/(n_{\elem{X},~\odot}/n_{\elem{H},~\odot}))$.
The absolute abundance of an element \elem{X} is $A(\elem{X}) = 12 + \log_{10}
(n_\elem{X}/n_\elem{H})$.
In \sectionname~\ref{sec:data} we present the astrometric and spectroscopic data
about the two stars relevant to the present discussion.
In \sectionname~\ref{sec:discussion} we discuss possible interpretations of the
abundance difference between the pair.
We summarize our discussions in \sectionname~\ref{sec:summary}.

\section{Data}
\label{sec:data}

\begin{deluxetable*}{c|cccc}
  \tablecaption{Astrometric and spectroscopic measurements of the pair
  \label{tab:kk}}
\tablehead{
  \colhead{}     & \colhead{}      & \sunanalog\         & \bizarreone\        & \colhead{} \\
  \colhead{Name} & \colhead{Units} & \colhead{HD 240429} & \colhead{HD 240430} & \colhead{Uncertainties}
}
\startdata
Sp Type                             &                & G0                     & G2                     &       \\
R.A.\tablenotemark{a}               & hh:mm:ss       & 23:51:55.21            & 23:52:09.42            &       \\
Dec.\tablenotemark{a}               & dd:mm:ss       & 59:42:48.16            &  59:42:26.08           &       \\
2MASS $J$\tablenotemark{a}          & mag            & $8.593 \pm 0.023$      & $8.415 \pm 0.026$      &       \\
$T_\mathrm{eff}$                    & K              & 5878                   & 5803                   & 25    \\
$\log{g}$                           &                & 4.43                   & 4.33                   & 0.028 \\
$v\sin{i}$                          & \kms\          & 1.1                    & 2.5                    &       \\
$[\elem{Fe}/\elem{H}]$              &                & 0.01                   & 0.20                   & 0.010 \\
Age\tablenotemark{b}                & Gyr            & $4.00_{-1.56}^{+1.51}$ & $4.28_{-1.03}^{+1.11}$ &       \\
$v_r$                               & \kms\          & $-21.2$                & $-21.2$                & 0.2   \\
$\varpi$\tablenotemark{a}          & mas            & $9.35 \pm 0.24$        & $9.41 \pm 0.25$        &       \\
$\mu_\alpha^*$\tablenotemark{a}    & mas\,yr$^{-1}$ & $89.25 \pm 0.66$       & $89.41 \pm 0.69$       &       \\
$\mu_\delta$\tablenotemark{a}      & mas\,yr$^{-1}$ & $-29.68 \pm 0.54$      & $-30.12 \pm 0.52$      &       \\
\hline
\multicolumn{5}{c}{$T_c < 1200$~K} \\
\hline
$A(\elem{Li})$\tablenotemark{c}     &                & $2.25$                 & $2.75$                 & 0.05      \\
$\elemH{C}$                         &                & $0.00$                 & $0.09$                 & 0.026 \\
$\elemH{N}$                         &                & $-0.06$                & $-0.01$                & 0.042 \\
$\elemH{O}$                         &                & $0.01$                 & $0.09$                 & 0.036 \\
$\elemH{Na}$                        &                & $-0.06$                & $-0.04$                & 0.014 \\
$\elemH{Mn}$                        &                & $-0.03$                & $0.00$                 & 0.020 \\
\hline
\multicolumn{5}{c}{$T_c > 1200$~K} \\
\hline
$\elemH{Mg}$                        &                & $0.01$                 & $0.19$                 & 0.012 \\
$\elemH{Al}$                        &                & $0.01$                 & $0.21$                 & 0.028 \\
$\elemH{Si}$                        &                & $0.00$                 & $0.16$                 & 0.008 \\
$\elemH{Ca}$                        &                & $0.02$                 & $0.23$                 & 0.014 \\
$\elemH{Ti}$                        &                & $0.02$                 & $0.20$                 & 0.012 \\
$\elemH{V}$                         &                & $0.02$                 & $0.20$                 & 0.034 \\
$\elemH{Cr}$                        &                & $0.01$                 & $0.17$                 & 0.014 \\
$\elemH{Fe}$                        &                & $0.01$                 & $0.20$                 & 0.010 \\
$\elemH{Ni}$                        &                & $-0.01$                & $0.21$                 & 0.014 \\
$\elemH{Y}$                         &                & $0.04$                 & $0.26$                 & 0.030 \\
\enddata
\tablenotetext{a}{From \tgas.}
\tablenotetext{b}{Derived in this work by isochrone fitting
  using the Yale-Yonsei model isochrones (\citealt{2013ApJ...776...87S}; see \sectionname~\ref{sub:ages}).}
\tablenotetext{c}{Absolute abundances from \citealt{jmlithium}.}
\tablecomments{
  All values are from \citealt{2016ApJS..225...32B} unless otherwise noted.
  The microturbulence parameter is fixed at $0.85$~\kms\ (\citealt{2015ApJ...805..126B}).
}
\end{deluxetable*}

\begin{figure}[htbp]
  \begin{center}
    \includegraphics[width=\linewidth]{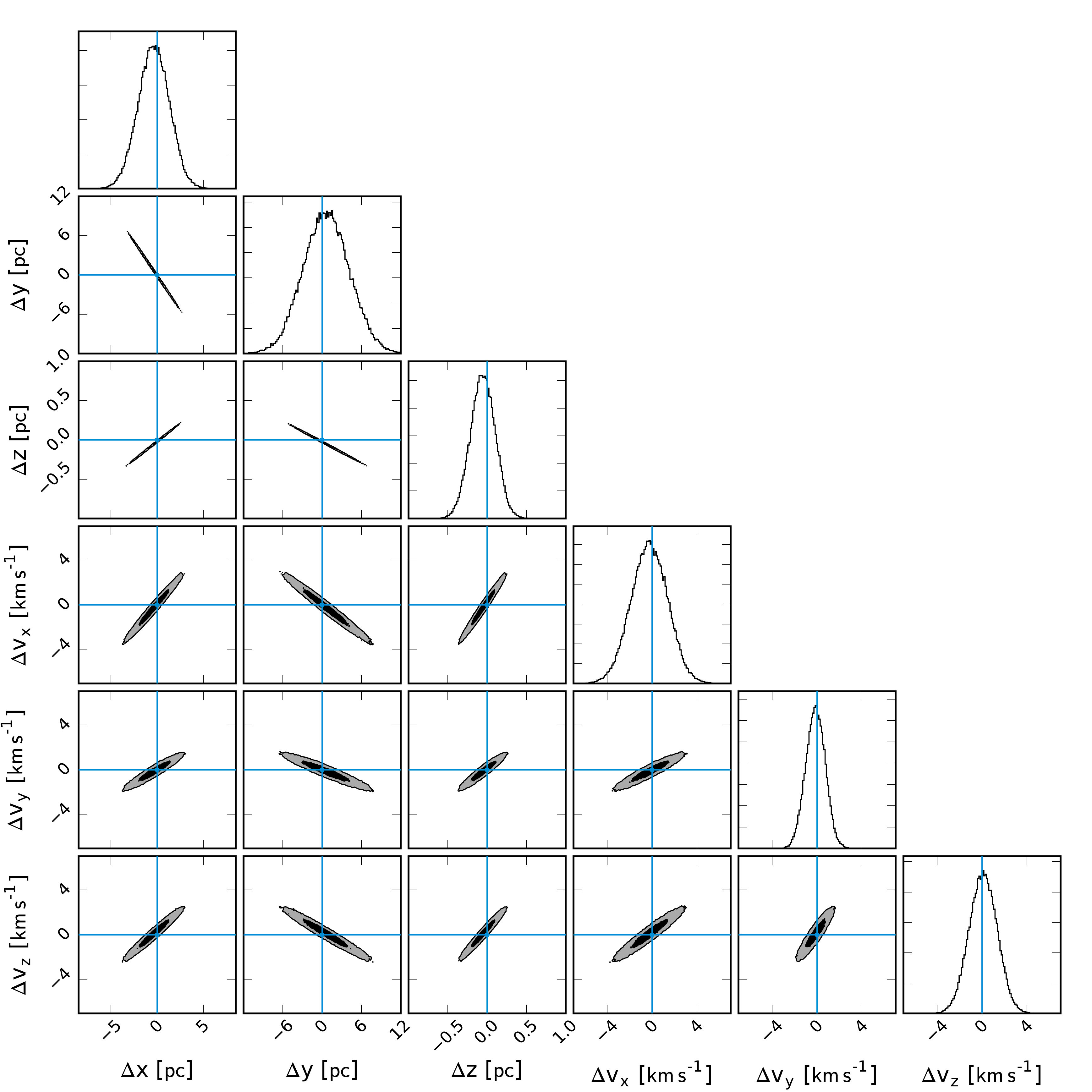}
  \end{center}
  \caption{%
    Differences in posterior samples over Galactocentric phase-space coordinates
    for the two stars \sunanalog\ and \bizarreone.
    \label{fig:dxdv}}
\end{figure}

\begin{figure}[htpb]
  \centering
  \includegraphics[width=0.9\linewidth]{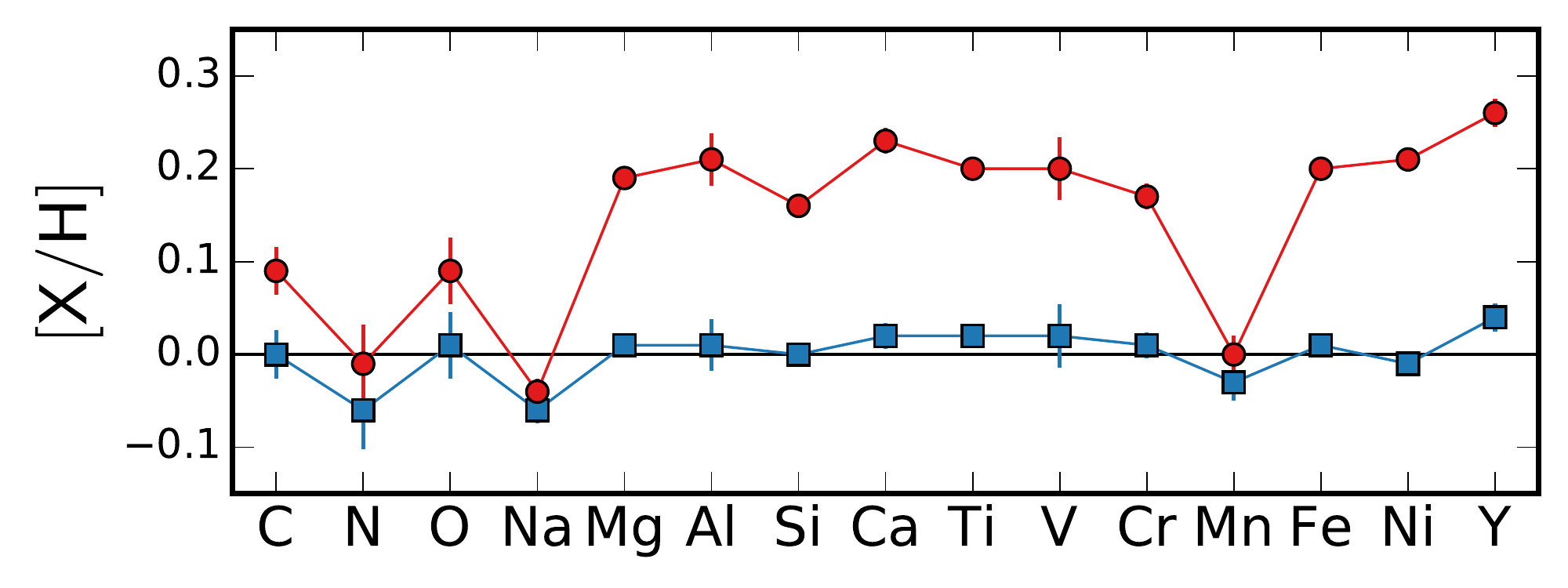}
  \caption{Abundances of the comoving pair, \sunanalog\ (blue) and \bizarreone\
    (red). Lines are drawn for each star only to guide the eye. \bizarreone\ is
    enhanced in \elem{Fe} by $\approx 0.2$~dex relative to \sunanalog\ along
    with \elem{Mg}, \elem{Al}, \elem{Si}, \elem{Ca}, \elem{Ti}, \elem{V},
    \elem{Cr}, \elem{Ni}, \elem{Y} yet not in \elem{C}, \elem{N}, \elem{O},
    \elem{Na}, and \elem{Mn}.
  }
  \label{fig:abundances}
\end{figure}

\begin{figure}[htpb]
  \centering
  \includegraphics[width=0.95\linewidth]{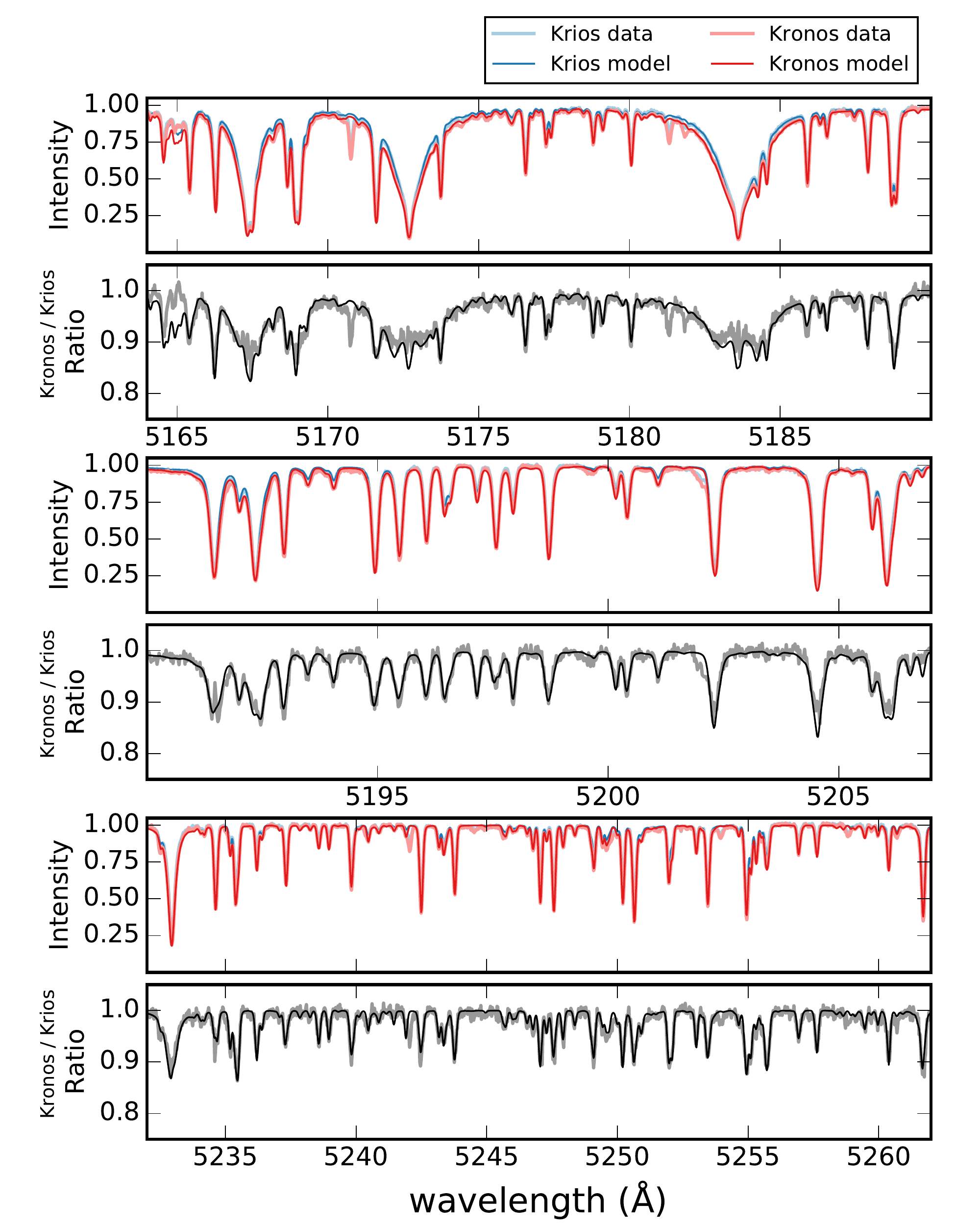}
  \caption{Selective segments of the spectra of \sunanalog\ and \bizarreone.
    Alternating sets of two rows show
    the continuum-normalized data and model in the upper panel,
    and the ratio (\bizarreone/\sunanalog) of data (gray) and model (black)
    in the lower panel.
  }
  \label{fig:spec1}
\end{figure}

\begin{figure}[htpb]
  \centering
  \includegraphics[width=0.95\linewidth]{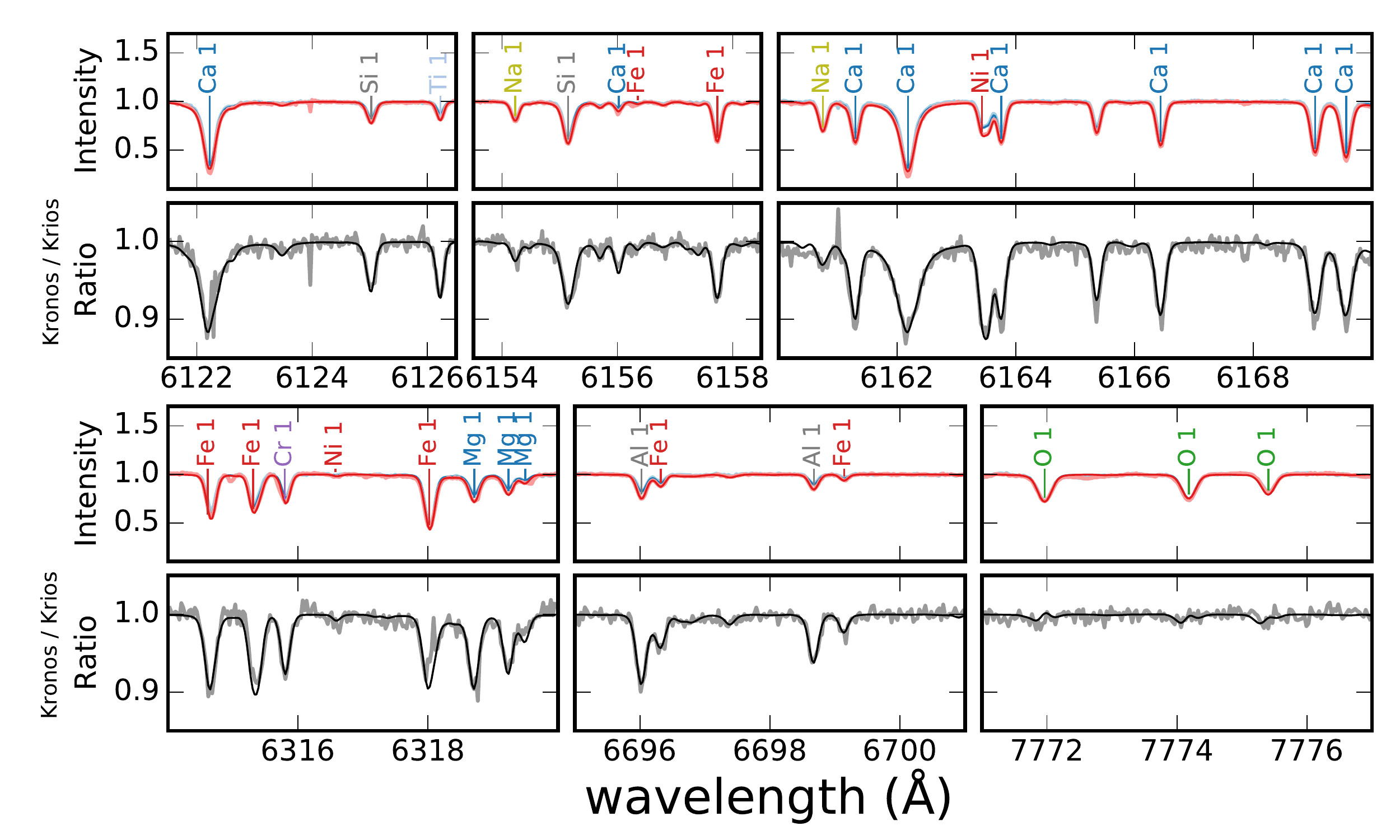}
  \caption{Same as \figname~\ref{fig:spec1}
    but for smaller portions of spectra at longer wavelengths that are
    not dominated by \elem{Fe}.
    We mark elements that give rise to strong absorption lines.
    Note that the lines of \elem{Na} and \elem{O}, which are under-enhanced
    in \bizarreone\ relative to \elem{Fe} or other refractory elements,
    show weaker residuals.
  }
  \label{fig:spec2}
\end{figure}

\begin{figure}[htpb]
  \centering
  \includegraphics[width=0.6\linewidth]{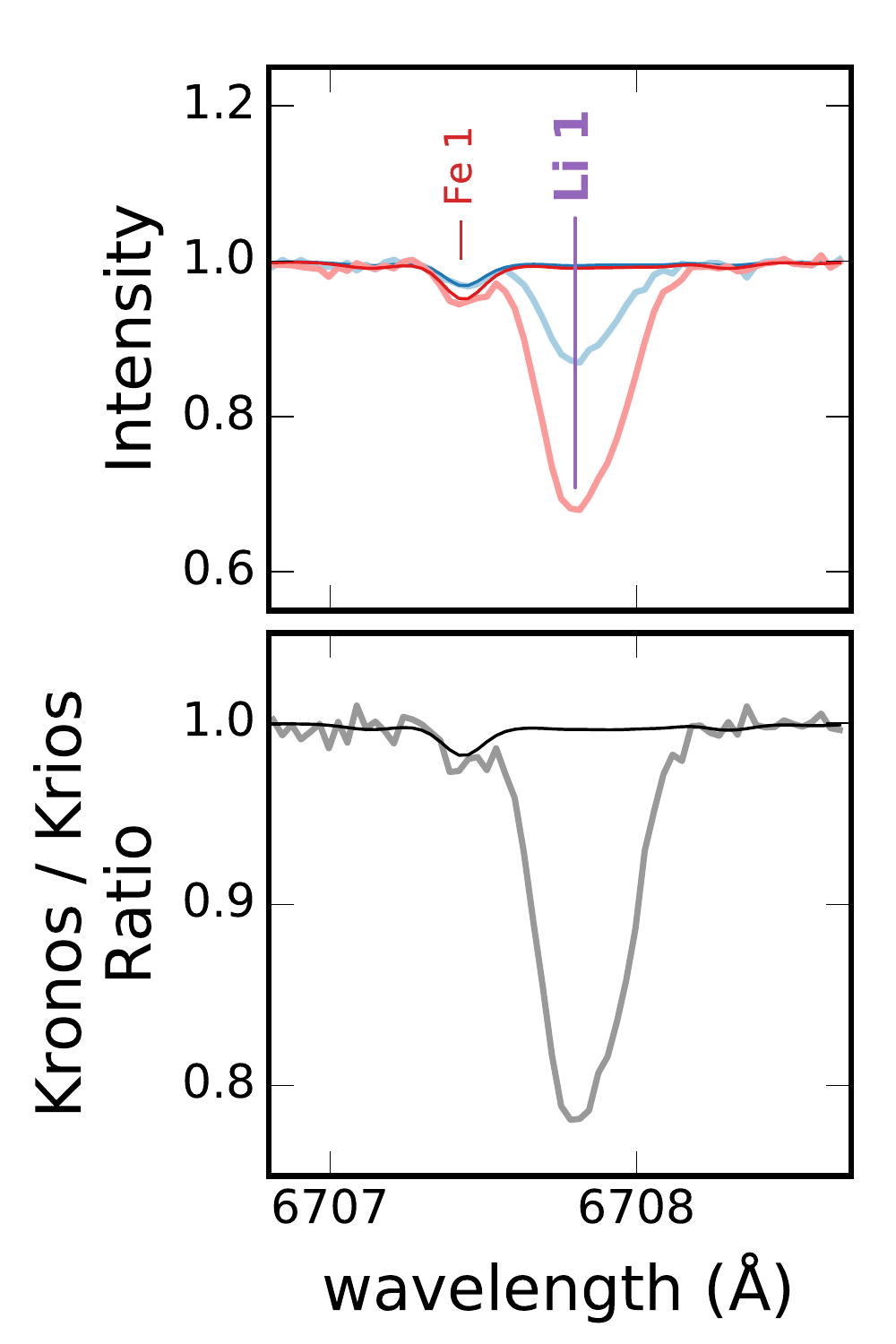}
  \caption{Lithium lines in the spectra of \bizarreone\ and \sunanalog.
    This line is studied in Myles \etal\ (in prep.).
    Line legends are the same as in \figname~\ref{fig:spec1}.
  }
  \label{fig:spec_lithium}
\end{figure}

\begin{figure}[htpb]
  \centering
  \includegraphics[width=0.95\linewidth]{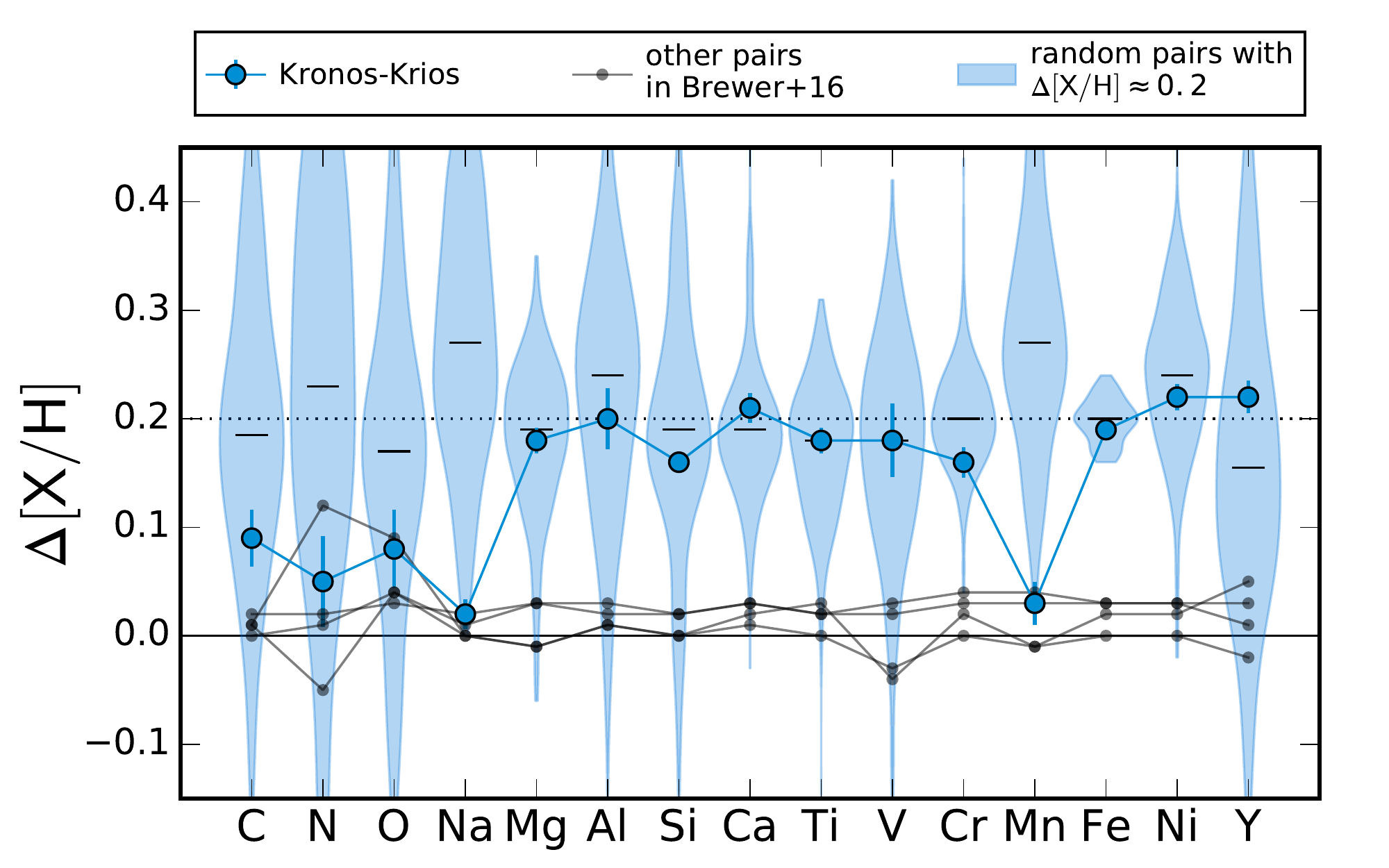}
  \caption{Abundance difference in this pair and other twin-like
    ($\Delta T_\mathrm{eff}\lesssim 100$~K) wide binaries in
    \citealt{2016ApJS..225...32B}.
    The differences in other pairs are small ($<0.05$~dex)
    for all elements except \elem{N} and \elem{O} which are the most
    uncertain, making the difference of $\approx 0.2$~dex seen in
    \bizarreone-\sunanalog\ rare.
    Additionally, we show the distribution of abundance differences
    between field stars with similar metallicity difference
    ($\Delta[\elem{Fe}/\elem{H}] \approx 0.2$)
    as violins with medians indicated by black line segments.
    These are random pairings of single stars in
    in \citealt{2016ApJS..225...32B} at two metallicity bins,
    $-0.025 < \feh < 0.025$ (160 stars) and $0.175 > \feh > 0.225$ (137 stars),
    similar to \bizarreone\ and \sunanalog.
    The difference is always taken to be
    $\mathrm{higher}\,\feh - \mathrm{lower}\,\feh$.
    Thus, the narrower range of $\Delta\feh$ is by construction.
    Random pairings of disk stars with similar $\Delta\feh$ usually show
    similar enhancement in all other elements
    unlike the pattern seen in \bizarreone-\sunanalog\ pair.
  }
  \label{fig:deltaXH}
\end{figure}

\sunanalog\ and \bizarreone\ were identified as a candidate comoving star pair
in our recent search for comoving stars using the proper motions and parallaxes
from the Tycho-Gaia Astrometric Solution catalog (\tgas), a component of \gaia\
\dr.
We refer the readers to this previous work (\citealt{2017AJ....153..257O}) for a
full explanation of the methodology behind this search and only include
a brief description here.
For a given pair, we compute the marginalized likelihood ratio between the
hypotheses (1) that a given pair of stars share the same 3D velocity vector,
and (2) that they have independent 3D velocity vectors, using only the
astrometric measurements from \tgas\ (parallaxes and proper motions).
We then select a sample of high-confidence comoving pairs by making a
conservative cut on this likelihood ratio.
In the resulting catalog of comoving pairs (\citealt{2017AJ....153..257O}),
the pair presented in this paper was assigned a group id of 1199,
and the marginalized likelihood ratio (Bayes factor)
between the two hypotheses is $\ln{\mathcal{L}_1/\mathcal{L}_2} = 8.52$,
well above the adopted cut value of 6.
The pair has also been previously recognized as a visual double star system
in Washington Double Star catalog (\citealt{2001AJ....122.3466M}).
We have checked that we do not find any possible additional comoving companions
by lowering the likelihood ratio cut for the stars around this pair.

In a separate effort to study detailed chemical abundances of potential
planet-hosting stars, high-resolution spectra of both stars were obtained using
the HIRES spectrograph on the Keck~I telescope, and analyzed
\citep{2016ApJS..225...32B}.
The spectral resolution is $R\approx 70000$ and the wavelength coverage is
$5164$--$7799$~\AA.
A typical signal-to-noise ratio in the spectral continuum is $>200$~per pixel.
The resulting measurements include elemental abundances for 15 chemical species
(C, N, O, Na, Mg, Al, Si, Ca, Ti, V, Cr, Mn, Fe, Ni, Y) as well as stellar parameters
and high precision radial velocities.
For the details of the spectral analysis, we refer the readers to
\citealt{2016ApJS..225...32B}.
Additionally, the \elem{Li} doublet at $6707.6$~\AA\
for this sample was investigated in a separate work (\citealt{jmlithium}).
We list all relevant astrometric and spectroscopic measurements including the
absolute abundances of \elem{Li} for the two stars in Table~\ref{tab:kk}.

The projected separation between the pair is 1.9\arcmin\ ($\approx 0.01$~pc),
and the 3D separation is $\approx 0.6$~pc.
Although selected based only on their astrometry, the two stars
have identical radial velocities within their uncertainties (Table~\ref{tab:kk}),
confirming that they are truly comoving.
Combining these precise radial velocities with the \gaia\ \tgas\ astrometry, we
can compare differences between the inferred 6D phase-space coordinates of the
two stars.
We start by generating posterior samples over the Heliocentric distance, $r$,
tangential velocities, $(v_{\alpha^*}, v_\delta)$, and radial velocity, $v_r$,
given the observed parallax, $\hat\pi$, proper motions,
$(\hat\mu_{\alpha^*}, \hat\mu_\delta)$, and radial velocity, $\hat v_r$
\footnote{$\alpha^*$ denotes the projection in right ascension direction,
  i.e., $\mu_{\alpha^*} = \dot\alpha \cos\delta$.
}.
We assume the noise is Gaussian, and the radial velocity measurements are
uncorrelated with the astrometric measurements.
If we define
\begin{eqnarray}
  \vec{\hat y} &=&
      \transp{\left(
        \begin{array}{c@{\hspace{1em}} c@{\hspace{1em}} c@{\hspace{1em}} c}
          \hat\pi &
          \hat\mu_{\alpha^*} &
          \hat\mu_\delta &
          \hat v_r
        \end{array}
      \right)}\\
  \vec{y} &=&
      \transp{\left(
        \begin{array}{c@{\hspace{1em}} c@{\hspace{1em}} c@{\hspace{1em}} c}
          r^{-1} &
          r^{-1}\,v_\alpha &
          r^{-1}\,v_\delta &
          v_r
        \end{array}
      \right)}
\end{eqnarray}
then the likelihood is
\begin{equation}
  \vec{\hat y} \sim \mathcal{N}(\vec{y}, \mat{C})
\end{equation} where $\mat{C}$ is the covariance matrix.
We adopt a uniform space density prior for the distance and an isotropic
Gaussian for any velocity component, $v$, with a dispersion $\sigma_v=25~\kms$
\begin{eqnarray}
p(r) &=&
  \begin{cases}
    \frac{3}{r_{\rm lim}^3} \, r^2 & \text{if } 0 < r < r_{\rm lim}\\
    0              & \text{otherwise}
  \end{cases}\\
p(v) &=& \frac{1}{\sqrt{2\pi}\,\sigma_v} \,
  \exp\left[-\frac{1}{2} \, \frac{v^2}{\sigma_v^2} \right] \quad .
\end{eqnarray}
For each of the two stars, we use \project{emcee}
(\citealt{2013PASP..125..306F}) to generate posterior samples in $(r, v_\alpha,
v_\delta, v_r)$ by running 64 walkers for 4608 steps and discarding the first
512 steps as the burn-in period.
For each sample, we convert the heliocentric phase-space coordinates into
Galactocentric coordinates assuming that the Sun's position and velocity are
$\vec x_\odot = (-8.3,\,0,\,0)~{\rm kpc}$ and $\vec v_\odot =
(-11.1,\,244,\,7.25)~\kms$ \citep[e.g.,][]{Schonrich:2010, Schonrich:2012}.

\figurename~\ref{fig:dxdv} shows differences in posterior samples converted to
Galactocentric phase-space coordinates for the two stars.
The differences in velocities are consistent with zero.
For a 2~\msun\ binary system, the Jacobi radius in the Solar neighborhood is
1.2~pc (\citealt{Jiang:2010aa}).
Thus, \bizarreone\ and \sunanalog\ are likely a bound system that formed coevally,
and we expect the two stars to have identical metallicities and abundance patterns.
However, one of the stars, \bizarreone\ is significantly more metal
rich than the other by 0.2~dex ($\approx 60\%$; \figname~\ref{fig:abundances}).
Moreover, not all elements are equally enhanced:
the abundances of \bizarreone\ show selective depletion in
\elem{C}, \elem{N}, \elem{O}, \elem{Na}, and \elem{Mn}
relative to \elem{Fe}.
\bizarreone\ also has a high surface \elem{Li} abundances, and the difference in
\elem{Li} abundance ($\approx 0.5$~dex) is the largest among all elements
measured.

The validity of the measured abundance differences is further demonstrated in
\figname~\ref{fig:spec1}, \ref{fig:spec2}, and \ref{fig:spec_lithium}, where we
show segments of the spectra and models of the two stars used to measure their
abundances (\citealt{2016ApJS..225...32B}).
As expected from their reported metallicity difference ($\Delta\feh \approx 0.2$),
the ratio of data and model between the two stars show significant
residuals for almost all metal line features, largely dominated by \elem{Fe}.
However, for lines of elements that are not as enhanced in \bizarreone\,
the residuals are much smaller in amplitude (\figname~\ref{fig:spec2}).
The \elem{Li} doublet, analyzed in a separate work (Myles \etal\ in prep.),
is clearly visible in the spectra of both stars, and is stronger in \bizarreone\
(\figname~\ref{fig:spec_lithium}).

We stress that none of the other four twin-like ($\Delta T_\mathrm{eff}
\lesssim 100$~K) wide binary pairs examined by \citealt{2016ApJS..225...32B}
show discrepancies in abundances between the stars at this level.
As shown in \figname~\ref{fig:deltaXH},
the differences in other pairs for all elements except \elem{N} and \elem{O},
which are also the most uncertain (\tablename~\ref{tab:kk}),
are less than $0.05$~dex, making \bizarreone-\sunanalog\ pair a significant outlier.
The statistical uncertainties for each parameter
presented in \tablename~\ref{tab:kk} from \citealt{2016ApJS..225...32B}
are estimated from repeated measurements of multiple spectra of the same stars.
We note that while there may be systematic uncertainties (bias) in the elemental
abundances of these two stars unconstrained by this procedure,
the systematic uncertainties, if any, for these two solar-type ``twin-like'' stars
with small differences in $T_\mathrm{eff}$ and $\log{g}$ are unlikely to wash out
the observed abundance differences of $\approx 0.2$~dex.

\begin{figure}[htbp]
  \begin{center}
    \includegraphics[width=\linewidth]{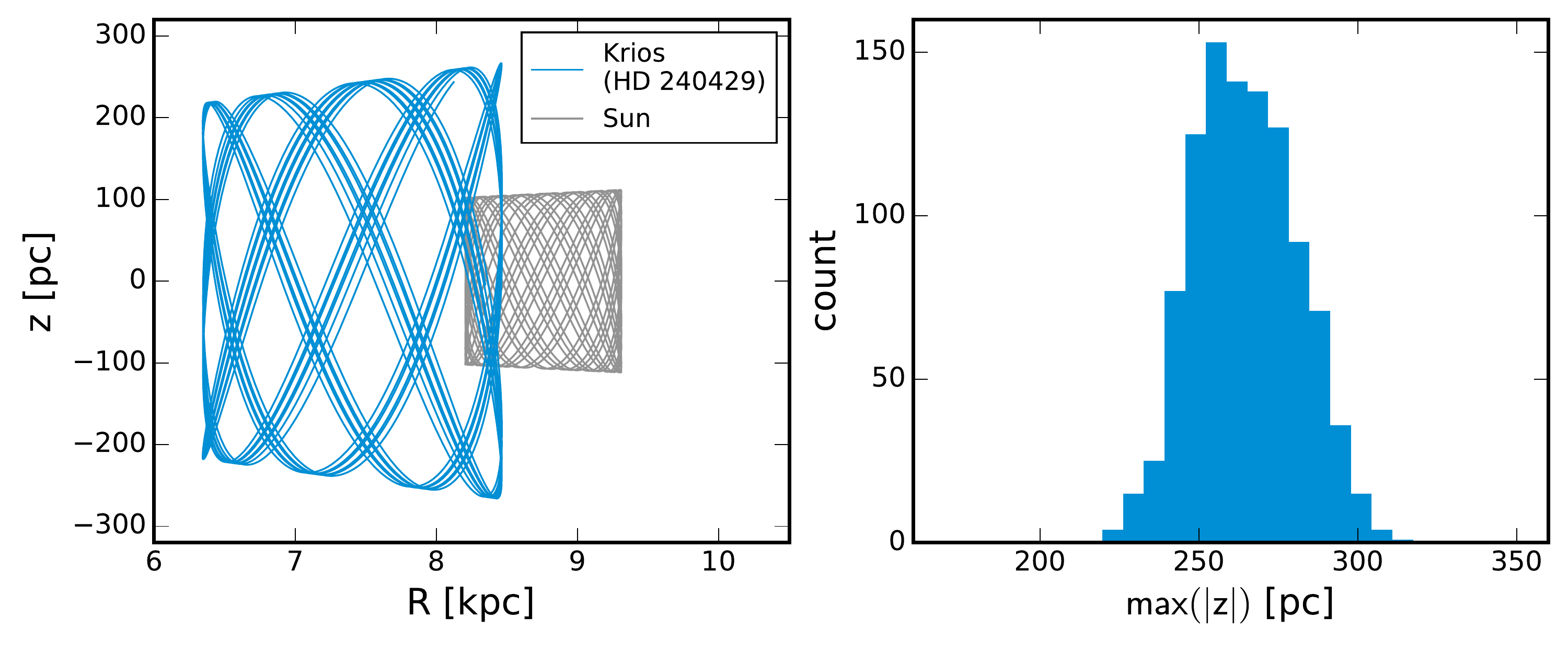}
  \end{center}
  \caption{Left panel: Galactic orbits computed for \sunanalog\ (black) and the
    Sun (grey).
    For \sunanalog, the initial conditions are set to the median of the
    posterior samples over the phase-space coordinates.
    The orbits are computed by integrating backwards from the present-day
    positions for $2.5$~Gyr with a time step of $0.5$~Myr using the Leapfrog
    integration scheme implemented in \project{Gala} (\citealt{gala}).
    Right panel: distribution of maximum $z$-heights for orbits computed from
    all posterior samples.
  }
  \label{fig:orbit}
\end{figure}

\begin{figure}[htpb]
  \centering
  \includegraphics[width=0.95\linewidth]{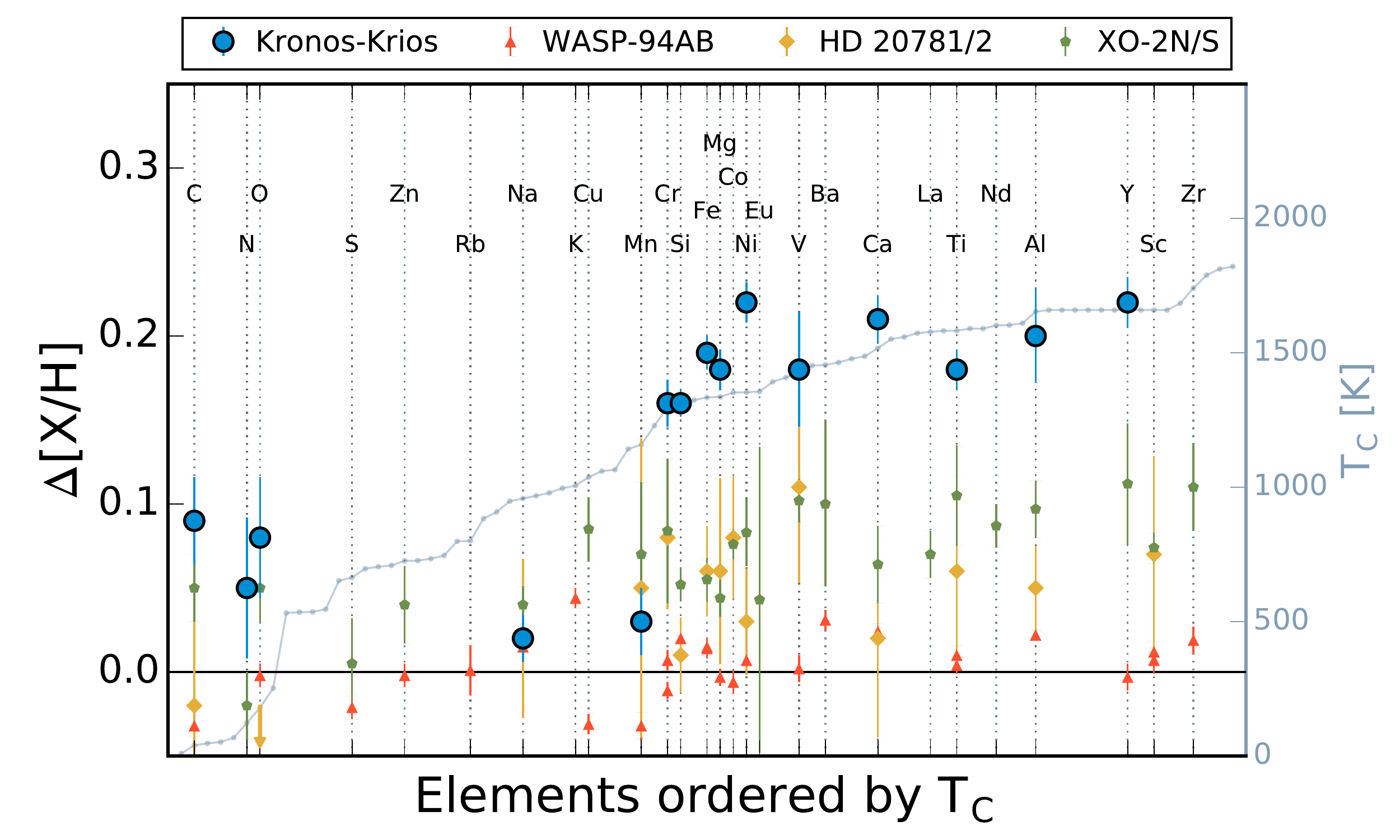}
  \caption{Abundance differences of the \bizarreone-\sunanalog\ pair
    ranked by the condensation temperature of elements for solar composition gas
    from \citealt{2003ApJ...591.1220L}.
    The condensation temperature may be read from the gray line and right y-axis.
    We show three wide binary systems selected from the literature:
    HD~20782/1 (\citealt{Mack:2014aa}, $\feh\approx0$),
    XO-2N/S (\citealt{Biazzo:2015aa}, $\feh\approx0.35$),
    and WASP-94AB (\citealt{Teske:2016aa}, $\feh\approx0.3$).
    Locations of elements with at least one measurement from any study
    are indicated by a vertical line and its symbol.
    Note that often multiple values are reported for one element corresponding
    to different ionization states in equivalent width analyses.
    No other pair studied so far were shown to have such large difference
    in metallicity or sharp contrast between (moderately) volatile and
    refractory elements as \bizarreone-\sunanalog.
  }
  \label{fig:relabun_tcrank}
\end{figure}

\section{Discussion}
\label{sec:discussion}

We discuss the possible origins of the peculiar abundance differences of
\bizarreone\ \& \sunanalog.
We first discuss the ages and coevality of the stars in this pair, and consider
both possibilities in which the two stars are or are not coeval.
Our favored scenario is discussed in the last subsection,
\sectionname~\ref{sub:accretion}.


\subsection{Stellar Ages \& Coevality}
\label{sub:ages}

Apart from their closeness in phase-space coordinates,
we can constrain the ages of the two stars
given the precise measurements of $\log(g)$ and $T_\mathrm{eff}$
by comparing these values to theoretical isochrones.
We use the distances (inferred from \gaia\ parallaxes), $V$-band magnitudes,
and $B-V$ colors to obtain bolometric luminosities of the two stars
(\citealt{2003AJ....126..778V}).
We then combine the luminosities with effective temperature, \elemH{Fe}, and
\elemH{Si} in order to interpolate the age, mass, and radius of each star using
a grid of Yale-Yonsei model isochrones (\citealt{2013ApJ...776...87S}).
The best-fit isochrone ages of \bizarreone\ and \sunanalog\ are
$4.28_{-1.03}^{+1.11}$~Gyr and $4.00_{-1.56}^{+1.51}$~Gyr, respectively,
consistent with them being coeval.

The surface lithium abundance in a Sun-like star decreases with its age due to
mixing induced by convection or rotation, which brings the lithium into the
interior ($T>2.5 \times 10^{6}$~K) where it will be destroyed by proton capture
burning.
In hotter stars with thin convective zones on the main sequence, most of this
mixing occurs in the pre-main sequence phase when the star is fully convective.
Thus, surface lithium abundance can be an indicator of stellar ages,
especially whether the star is very young ($\lesssim 1$~Gyr).
The absolute $\elem{Li}$ abundance of solar-type stars also correlates steeply
with the effective temperature (e.g., \citealt{Chen2001,2012ApJ...756...46R}).
Generally, cooler stars with larger convective envelope have lower \elem{Li}
abundances.
The absolute $\elem{Li}$ abundance of $2.25$~dex for \sunanalog\
is typical for its \teff.
On the other hand, the lithium abundance ($A(\elem{Li}) = 2.75$) of \bizarreone,
which has lower \teff\ than \sunanalog, is not only higher than that of
\sunanalog\ but also much higher compared to other field stars of similar \teff.
Given the overall higher metal abundances and the peculiar abundance patterns
in \bizarreone, it is unclear, however, whether this higher $\elem{Li}$
abundance means a younger age or something else.
For example, \citealt{Casey:2016aa} attributes the presence of $\elem{Li}$-rich
red giant stars to the engulfment of substellar companions such as gas giant
planets or brown dwarfs which may replenish $\elem{Li}$.

The surface lithium abundance of \bizarreone\ is the only indicator of a
younger age.
If the two stars were only several hundred Myrs old, then
they may have been part of a larger comoving group of stars.
However, as we mention above (\sectionname~\ref{sec:data}), there is no
evidence in our search of comoving pairs using \tgas\ that the two stars belong
to a larger group of young stars.
Very young stars often show signs of activity such as
X-ray emission from magnetic activity, emission lines, or infrared excess due to
circumstellar disks (\citealt{1999ARA&A..37..363F,1987ApJ...312..788A}).
We have compiled GALEX, Tycho-2, 2MASS, and WISE photometry for these stars,
and found no evidence for indications of activity in their spectral energy
distributions.
The low $v\sin(i)$ values (\tablename~\ref{tab:kk}) also argue against very
young ages that would be inferred from the surface lithium abundance.
Finally, we computed the Galactic orbit of the pair using the median of the
posterior sample over the phase-space coordinates of \sunanalog, in a Milky
Way-like gravitational potential (similar to \texttt{MWPotential2014} from
\citealt{Bovy:2015}) using \project{Gala} (\citealt{gala}).
The pair's fiducial orbit has a vertical action larger than the Sun, favoring
an older age (\citealt{Wielen:1977,Aumer:2016}).
We therefore conclude that the two stars are most likely coeval, $\sim 4$~Gyr
old main sequence stars, and that the unusually high \elem{Li} abundance of
\bizarreone\ requires an alternative explanation.

\subsection{Exchange Scattering}
\label{sub:exchange_scattering}

While the data described above strongly suggests that the two stars are coeval,
this subsection explores the possibility that this pair is still not a
primordial binary.
Two stars unrelated at birth may end up in a binary system via a binary-single
scattering event that results in an exchange of binary members.
In order to estimate the rate at which any binary-single
event will produce a wide binary system such as \sunanalog\ and \bizarreone,
we may consider the rate at which this wide binary will scatter with a field star to
result in an exchange reaction.
The cross-section of exchange scattering for a binary with semi-major axis $a$ is
\begin{eqnarray}
  \sigma_\mathrm{ex} = \frac{640}{81} \pi a^{2} \left(\frac{v_i}{v_c}\right)^{-6}
  \label{eq:crosssection}
\end{eqnarray}
where $v_i$ is the incoming velocity, and $v_c$ is the critical velocity,
defined as
\begin{eqnarray}
  v_c^2 = G \frac{m_1 m_2 (m_1 + m_2 + m_3)}{m_3 (m_1 + m_2)} \frac{1}{a}\,\,.
\end{eqnarray}
\eqname~\ref{eq:crosssection} is appropriate when $v_i/v_c \gg 1$
(\citealt{Hut:1983aa,Hut:1983ab}), which is the case for wide binaries
scattering with field (disk) stars.
If we assume that field stars are made of solar-mass stars with a constant
number density $n=1$~pc$^{-3}$, and the incoming velocity of field stars is
$10$~km\,s$^{-1}$, the rate of exchange scattering is
\begin{eqnarray}
  n \sigma_\mathrm{ex} v_i = 6.82\times 10^{-8}\,\mathrm{Gyr}^{-1}
  \frac{n}{\mathrm{pc}^{-3}} \frac{\mathrm{pc}}{a}
  \left(\frac{10~\mathrm{km}\,\mathrm{s}^{-1}}{v_i}\right)^5\,,
\end{eqnarray}
low enough to be negligible.

An exchange scattering scenario is unlikely to be able to explain
the observed abundance difference pattern of \bizarreone\ and \sunanalog.
We test this by randomly drawing pairs of stars in the sample of
\citealt{2016ApJS..225...32B} from two \feh\ bins at $\feh = 0 \pm 0.025$
and $\feh = 0.2 \pm 0.025$, each similar to \sunanalog\ and \bizarreone.
In \figname~\ref{fig:deltaXH}, we compare the observed abundance difference of
\bizarreone-\sunanalog\ with the distribution of abundance differences
from 300 random pairs.
We see that when a star is enhanced in $\elem{Fe}$ by $0.2$~dex,
all other elements are typically enhanced at a similar level, with some variations.
Specifically, for a typical star with $\feh \approx 0.2$~dex, we generally
expect $[\elem{Na}/\elem{Fe}] > 0$ and $[\elem{Mn}/\elem{Fe}] > -0.1$
(\citealt{Battistini:2015aa,Bensby:2003aa}) making the low
[\elem{Na}/\elem{Fe}] and [\elem{Mn}/\elem{Fe}] seen in \bizarreone\ very
unlikely to arise from variations in Galactic chemical evolution.

\subsection{Chemical Inhomogeneity in Star Formation}
\label{sub:chemical_inhomogeneity_in_star_formation}

In this subsection, we explore the hypothesis that chemical inhomogeneity
within the birth cloud is the source of the observed abundance difference.
There is ample evidence against this scenario as most wide binaries show a
difference in \feh\ less than $0.02$~dex
(\citealt{Desidera:2004aa,Gratton:2001aa}).
Even when a significant difference is detected with high-precision abundance
measurements, the difference is typically $\sim 0.05$~dex (see
\figname~\ref{fig:relabun_tcrank} and \sectionname~\ref{app:review}).
Consistent with these results, none of the other seven similar wide binaries
examined in \citealt{2016ApJS..225...32B} show such large differences in
abundances though there is generally a larger spread in $\elem{C}$, $\elem{N}$
and $\elem{O}$, and some pairs show a difference in particular elements as
large as $\approx 0.15$~dex.
The median and maximum \feh\ difference between component stars in the other
seven pairs is $0.02$~dex and $0.09$~dex, respectively.
The differences are even smaller (maximum $\Delta\feh = 0.03$~dex) if we
compare only twin-like ($\Delta T_\mathrm{eff} \lesssim 100$~K) pairs
(\figname~\ref{fig:deltaXH}, black lines).
Thus, a difference of $\approx 0.2$~dex seen in \bizarreone-\sunanalog\ pair is
unlikely to be due to chemical inhomogeneity in the birth cloud.

\begin{figure}[htpb]
  \centering
  \includegraphics[width=0.95\linewidth]{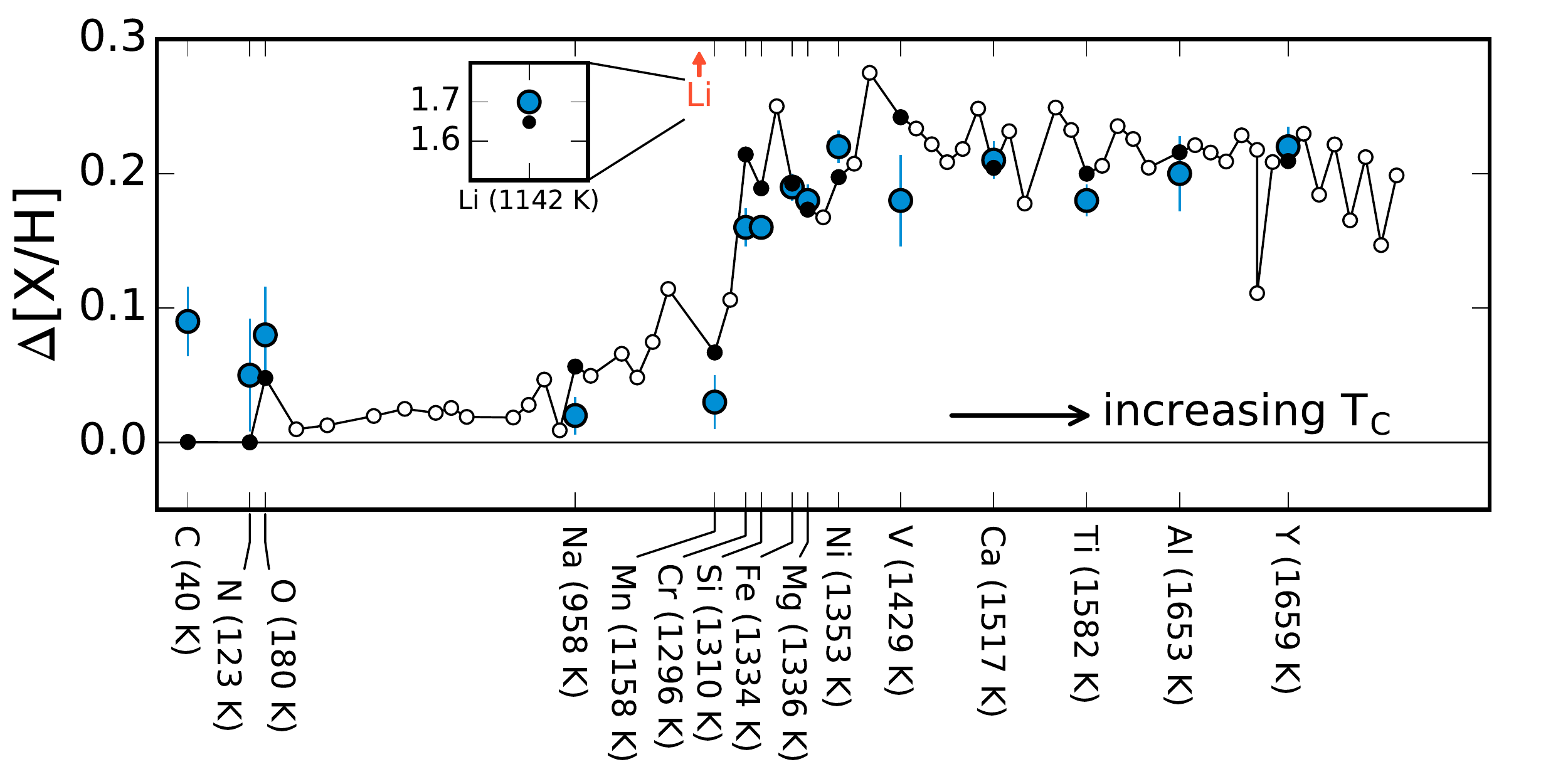}
  \caption{
    Comparing the observed abundance difference ($\bizarreone-\sunanalog$; blue
    circles) to the expected change in solar surface abundance after adding
    \maccreted\ of material with bulk Earth composition
    (\citealt{2003TrGeo...2..547M}; black open and filled circles).
    The assumed mass fraction in the convective zone is $0.02$.
    All astronomical metals are ordered by their \Tcondens\ for solar
    composition gas on the $x$-axis.
    For the predictions, we highlight elements measured for \bizarreone-\sunanalog\ pair
    in filled circles, while those without a measurement are left open.
    The close match with the observed abundance difference in \bizarreone-\sunanalog\ pair
    suggests that the abundance difference may be due to accretion of
    $15$~\mearth\ of rocky planetary material.
    The element \elem{Li} is off the plot and indicated in the inset.
  }
  \label{fig:toycalc}
\end{figure}

\subsection{Accretion of rocky planetary material}
\label{sub:accretion}

Another possibility that two coeval stars may end up with different surface
abundances is accretion of planetary material after birth.
In a multi-planet system, dynamical instabilities triggered by planet-planet
scattering (\citealt{1996Sci...274..954R,1996Natur.384..619W}) or encounters with a
field star (\citealt{Malmberg:2011aa}) can lead to planet ejection or accretion.
Indeed, it is an important goal of many exoplanet studies
to detect chemical signatures of planet formation or accretion,
distinguish them from Galactic chemical evolution, and
connect them to theories of evolution of planetary systems.
One approach that is free from confusion with Galactic chemical evolution
is to compare two almost identical stars in a wide binary system.
Assuming that the component stars were born together with identical
initial composition, we may see a difference in their surface abundances
if the two stars then accreted different amounts of planetary material.
The resulting abundance difference may depend on the condensation
temperatures of elements in the protoplanetary disks from which the accreted
planets formed, as their compositions depend on the radial temperature gradient
in the disk.

In Figure~\ref{fig:relabun_tcrank}, we show the abundance difference
between \bizarreone\ and \sunanalog\ ordered by the rank of \Tcondens\
of each element.
The equilibrium condensation temperatures for the composition of solar system
are taken from \citealt{2003ApJ...591.1220L} (Table~8).
The difference seen in \bizarreone-\sunanalog\ is
compared to HD~20781/2, XO-2N/S, WASP-94A/B in \figname~\ref{fig:relabun_tcrank}.
The metallicity difference of $\approx 0.2$~dex observed in this pair
is larger than the differences seen in any other pairs studied so far
(see also Appendix~\ref{app:review}).
The five under-enhanced elements in \bizarreone\
relative to \sunanalog\ are the five most volatile in all elements measured.
The difference in \elem{Mn} ($\Tcondens = 1158$~K) and
\elem{Cr} ($\Tcondens = 1296$~K) suggests a break in $\Tcondens \approx 1200$~K.
This $\Tcondens$-dependent trend of $\Delta\elemH{X}$,
combined with the enhanced $\elem{Li}$ abundance ($A(\elem{Li}) = 2.75$),
strongly suggests that accretion of rocky material has occurred in \bizarreone.

How much mass of rocky material is needed to explain an increment of
$\approx 0.2$~dex?
We carry out simple toy calculations of the expected $\Delta\elemH{X}$
in a Sun-like star's atmosphere by adding a certain mass of bulk Earth composition
under these simplifying assumptions:
\begin{itemize}
  \item The material added is instantly and completely mixed through the star's convective zone.
  \item The atmospheric composition that we measure is identical throughout
    the star's radiative and convective zone.
  \item The surface abundance of the star has been altered only by the
    accretion event(s).
\end{itemize}

We take the solar abundances, $\elemH{X}$, of element \elem{X}
(\citealt{Asplund:2009aa}) which can be converted to mass fraction as
\begin{equation}
  f_{X,\mathrm{photo}} = \frac{10^{\elemH{X}} \, m_X}{\Sigma_X 10^{\elemH{X}} \, m_X}
\end{equation}
where $m_X$ is the mass of each element in, e.g., atomic mass unit.
Assuming that the accreted material has a total mass $M_\mathrm{acc}$, and the
mass fraction in each element $f_{X,\mathrm{acc}}$,
the abundance difference is
\begin{equation}
  \Delta\elemH{X} = \log_{10} \frac{f_{X,\mathrm{photo}}\,f_\mathrm{CZ}\,M_\mathrm{star}
    + f_{X,\mathrm{acc}}\,M_\mathrm{acc}}
    {f_{X,\mathrm{photo}}\,f_\mathrm{CZ}\,M_\mathrm{star}}
\end{equation}
where $f_\mathrm{CZ}$ is the fraction of the star's mass in the convective envelope.
We assume $f_\mathrm{CZ} = 0.02$ (\citealt{2013ApJ...776...87S}),
and take the composition of bulk Earth from a chondritic model of the Earth
(\citealt{2003TrGeo...2..547M}).
Similar calculations have been performed by, e.g., \citet{Chambers:2010aa},
\citet{Mack:2014aa,Mack:2016aa}.

Figure~\ref{fig:toycalc} shows the expected change of surface abundances of
metals in a Sun-like star after \maccreted\ of material with composition of
bulk Earth is added.
A volatility trend such that more volatile (low \Tcondens) elements are more
depleted in the Earth relative to CI or other carbonaceous chondrites
has long been known (\citealt{mcdonough2001composition}).
This trend is presumed to be closely related to the formation of terrestrial
planets and, in particular, to the radial temperature gradient in a
protoplanetary disk.
The trend resulting from adding \maccreted\ of bulk Earth
provides an overall good match to the observed $\Delta\elemH{X}$,
suggesting that the refractory-enhanced star, \bizarreone\,
accreted \maccreted\ more of rocky planetary material than \sunanalog.

What about \elem{Li}?
The element \elem{Li} is worth special attention in the context of the
accretion scenario.
Because Li is present in either carbonaceous chondrites or bulk Earth with a
concentration of $1-1.5$~ppm in mass (\citealt{2003TrGeo...2..547M}), but is
depleted quickly within the first Gyr on the surface of a Sun-like star
(\citealt{Thevenin2017,Baraffe2017}), accretion of either material at later
times will significantly replenish the lithium on the star's surface.
For the present-day Sun ($A(\elem{Li}) = 1.05$), the accretion of \maccreted\ of
bulk Earth-like material would result in $\Delta\elemH{Li} \approx 1.65$~dex
(see the inset of \figname~\ref{fig:toycalc}).
This closely matches what we find: the \elem{Li} abundance of \bizarreone\ is
$A(\elem{Li}) = 2.75$ (Table~\ref{tab:kk}, \citealt{jmlithium})
approximately $1.7$~dex higher than the solar value.

We stress that while the calculation carried out is useful in
an order-of-magnitude sense, further investigation of each of the simplifying
assumptions made is warranted.
In addition, the composition of bulk Earth has some uncertainties.
For example, the reported bulk Earth concentration of the siderophile element
\elem{Mn}, varies from $800$ to $\approx 2000$~ppm
(\citealt{1998psc..book.....L,mcdonough2001composition,2003TrGeo...2..547M})
mainly due to the uncertainty of the Earth's core composition.
Given these limitations, the level of agreement for $\Delta\elemH{X}$ {\it and}
\elem{Li} for \bizarreone\ is remarkable.

The fractional mass in the convective zone of solar-type stars
decrease dramatically in the first Gyr,
and then stays nearly constant at $\approx 2$~\% (\citealt{2013ApJ...776...87S}).
Because the accreted mass $M_\mathrm{acc}$ is proportional to $f_\mathrm{CZ}$,
given the large metallicity enhancement ($\approx 0.2$~dex),
the accretion must have happened after a thin convective envelop is established.
Otherwise, the accreted mass would be unreasonably high.
Thus, it is plausible that a dynamical process after the planet formation ended
is responsible for pushing rocky planets in.

Finally, we mention that detection of $^6 \elem{Li}$ provides a strong test for
this scenario.
This isotope of \elem{Li} is destroyed at even lower temperatures than
$^7\elem{Li}$, and theoretically expected to be absent
(\citealt{1997ARA&A..35..557P}).
However, an accretion of rocky material could have replenished $^6 \elem{Li}$.
Because $^6\elem{Li}$ lines are slightly longer in wavelengths,
presence of $^6\elem{Li}$ increases the asymmetry of \elem{Li} $6707.6$ feature.
Depending on how recent the accretion was and how fast $^6 \elem{Li}$ is
depleted on the main sequence, this feature may be detectable.
This is a very subtle effect that requires a higher signal-to-noise, higher
resolution spectra and careful modelling effort
(see e.g., \citealt{Israelian:2001,2002MNRAS.335.1005R}).
Such investigation was not warranted by the current data (\citealt{jmlithium}).

\section{Summary}
\label{sec:summary}

We report and discuss the discovery of a comoving pair of bright
solar-type stars HD~240430 and HD~240429 (G0 and G2) with very different
metallicities ($\Delta\feh \approx 0.2$~dex), and condensation temperature
(\Tcondens)-dependent abundance differences.
The more metal-rich of the two stars, HD~240430 (\bizarreone), shows enhancement in
all ten elements with $\Tcondens > 1200$~K including \elem{Fe}, while
under-enhanced in the five elements, \elem{C}, \elem{N}, \elem{O}, \elem{Na}, and
\elem{Mn} with $\Tcondens < 1200$~K relative to HD~240429 (\sunanalog).
It also has an anomalously high surface \elem{Li} abundance for its age of
$\sim 4$~Gyr, and its effective temperature very close to that of the Sun.
We consider that the comoving pair may have formed from two stars of different
birth origins in an exchange scattering event
(\sectionname~\ref{sub:exchange_scattering}), or that there may be chemical
inhomogeneity in the birth cloud
(\sectionname~\ref{sub:chemical_inhomogeneity_in_star_formation}) to find
both unlikely.

In order to explain the $\Tcondens$-dependent enhancement and high \elem{Li}
abundance of \bizarreone, we consider the accretion of planetary material as the
most plausible cause (\sectionname~\ref{sub:accretion}).
We argue that an accretion of \maccreted\ of bulk Earth composition to
\bizarreone\ after its thin convective zone is in place can explain the
enhancement in both refractory elements and lithium.
What triggered the planet engulfment in the two comoving stars remains unclear.
One possibility is that a fly-by interaction with a field star could have
triggered eccentricity excitation of outer planets, which may
have propagated inward through planet-planet scattering, leading to the
accretion of inner rocky planets (\citealt{Zakamska:2004aa,Malmberg:2011aa}).
If this is the case, there may be surviving, highly-eccentric giant planets
potentially detectable with future data releases of the \gaia\ mission.

The two stars have not been included in any publicly released data from planet
search programs.
We have begun a precision radial velocity campaign for the two stars and early
indications are that there are no close in giant planets.
If both stars have accreted planetary material, it would be very interesting to
search for the existence and architectures of the planetary systems left
behind.

\acknowledgements
We thank Andy Casey for bringing $^{6}\elem{Li}$ into our attention.
We thank Megan Bedell and Andy Casey for valuable discussions,
and Keith Hawkins, Nathan Leigh, and Josh Winn for comments
on the early version of the draft.
The Flatiron Institute is supported by the Simons Foundation.

This work has made use of data from the European Space Agency (ESA) mission
{\it Gaia} (\url{http://www.cosmos.esa.int/gaia}), processed by the {\it Gaia}
Data Processing and Analysis Consortium (DPAC,
\url{http://www.cosmos.esa.int/web/gaia/dpac/consortium}). Funding for the DPAC
has been provided by national institutions, in particular the institutions
participating in the {\it Gaia} Multilateral Agreement.
This publication makes use of data products from the Two Micron All Sky Survey,
which is a joint project of the University of Massachusetts and the Infrared
Processing and Analysis Center/California Institute of Technology, funded by
the National Aeronautics and Space Administration and the National Science
Foundation.
This publication makes use of data products from the Wide-field Infrared Survey
Explorer, which is a joint project of the University of California, Los
Angeles, and the Jet Propulsion Laboratory/California Institute of Technology,
funded by the National Aeronautics and Space Administration.

\software{
  This research utilized:
  \texttt{Astropy} (\citealt{Astropy-Collaboration:2013}),
  \texttt{corner.py} (\citealt{corner}),
  \texttt{emcee} (\citealt{2013PASP..125..306F}),
  \texttt{IPython} (\citealt{Perez:2007}),
  \texttt{matplotlib} (\citealt{Hunter:2007}),
  \texttt{numpy} (\citealt{Van-der-Walt:2011}),
  and \texttt{pandas} (\citealt{pandas}).
}

\appendix
\section{
  Review of Detailed Chemical Abundance Studies of Stars in Comoving Pairs}
\label{app:review}

We review and summarize a handful of
wide binary systems that have been studied in their detailed chemical
abundances so far with high-resolution spectroscopy.
These systems are
  16 Cygni A/B,
  HD~20782/HD~20781,
  HD~80606/HD~80607,
  XO-2N/XO-2S,
  HAT-P-1,
  WASP-94A/WASP94-B, and
  HD~133131A/HD~133131B.
We focus on key characteristics of stars and planets, and interpretations of
any trend in $\Delta\elemH{X}$ with \Tcondens.
Interested readers may also consult \citealt{2016arXiv161104064M}.

{\bf 16 Cygni A/B:}
The chemical composition of this well know pair of solar-type stars (G1.5/G3)
has been studied many times.
The hotter star 16 Cyg A has no detected planets, but has an M dwarf companion
$\sim 70$~AU away in projected separation which is probably physically
associated (\citealt{2002ApJ...581..654P}), and may have affected planet
formation process around the star
(\citealt{1996ApJ...458..312J,2005MNRAS.363..641M}).
The other star, 16 Cyg B, hosts a giant planet on an eccentric orbit ($e=0.63$,
\citealt{1997ApJ...483..457C}).
While past measurements of metallicity and abundance difference between the two
stars reported conflicting results
(\citealt{2001ApJ...553..405L,2011ApJ...737L..32S}), recent studies using high
quality spectra (\citealt{2011ApJ...740...76R,2014ApJ...790L..25T})
consistently reported that A is more metal rich than B by $\approx 0.04 \pm
0.005$~dex.
However, there is still a disagreement between studies on
whether abundance differences shows a correlation with $\Tcondens$ as well as
its interpretation.
\citealt{2014ApJ...790L..25T} suggested that formation of $1.5-6$~\mearth\
rocky core for the giant planet around 16 Cyg B can explain the offset and the
positive correlation between $\Delta\elemH{X} (\mathrm{A}-\mathrm{B})$
and \Tcondens. On the other hand, \citealt{2011ApJ...740...76R},
who found no correlation, argued that forming giant planets
results in an overall shift in all elements.

{\bf HD~20782/HD~20781:}
Two common proper motion G dwarf stars (G2/G9.5) with a projected separation of
$\sim9000$~AU (corresponding to 4.2\arcmin\ sky separation) and solar metallicity
host close-in giant planets.
HD~20782 hosts a Jupiter-mass planet on a very eccentric ($e\approx 0.97$)
orbit with a pericenter distance of 1.4~AU while HD~20781 hosts two
Neptune-mass planets within 0.3~AU with moderately high eccentricity
($e\sim0.1-0.3$).\footnote{
  The two stars were monitored by \project{HARPS} campaign, and it has recently
  been reported by \citealt{2017arXiv170505153U} that HD~20781 hosts four
  planets between $M\sin(i)\approx 0.006-0.04$~\mjupiter\ with $e \le 0.11$
  within $\approx 0.35$~AU.}
The measured abundances of 15 elements between the two stars are consistent
with each other (\citealt{Mack:2014aa}).
However, \citealt{Mack:2014aa} argued that there is a moderately significant
($\sim 2\sigma$) positive slope of $\approx 10^{-5}$~dex\,K$^{-1}$ with
increasing \Tcondens\ for $\Tcondens>900$~K elements (namely, Na, Mn, Cr, Si,
Fe, Mg, Co, Ni, V, Ca, Ti, Al, Sc leaving out C and O of their measurements) in
the abundances of each star {\it individually}.
They suggest that this slope is evidence that the stars accreted
$10-20$~\mearth\ of \elem{H}-depleted rocky material during giant planet
migration.

{\bf HAT-P-1:}
This pair of G0 stars separated by 11\arcsec\ with $\feh\approx0.15$ has
different planetary systems:
the secondary star is known to host one transiting giant planet
while no planet has been discovered around the primary star.
The two stars are identical
in metallicities and abundances for 23 elements measured with
the mean error of $0.013$~dex (\citealt{Liu:2014aa}).
Thus, it seems that the presence of close-in giant planet does not necessarily
lead to atmospheric pollution of its host star.

{\bf HD~80606/HD~80607:}
Similar to HAT-P-1, no significant chemical difference is found between two
common proper motion G5 stars with super-solar metallicity ($\feh \approx
0.35$). HD~80606 which hosts a very eccentric ($e\approx0.94$) giant planet and
HD~80607 which has no detected planets (\citealt{Saffe:2015aa,Mack:2016aa}).

{\bf XO-2N/XO-2S:}
A few independent studies have investigated this pair of G9 stars with
super-solar metallicity ($\feh \gtrsim 0.35$).
XO-2N hosts a giant planet while XO-2S is known to host two giant planets with
masses $0.26 \mjupiter$ and $1.37 \mjupiter$ on moderately eccentric ($\approx
0.15$) orbits at $<0.5$~AU.
A significant difference of metallicity ($\gtrsim0.05$~dex) is detected between
the two stars with a possible correlation with \Tcondens\
(\citealt{Ramirez:2015aa,Biazzo:2015aa} although see also
\citealt{Teske:2015aa,Teske:2013aa}).
At low \Tcondens, the difference $(\mathrm{N}-\mathrm{S})$ in volatile elements
differ by $\sim 0.01$~dex while the range of difference spans upto $0.1$~dex at
$\Tcondens>1600$~K.

\citealt{Ramirez:2015aa} suggested that the small overall depletion
($\approx 0.015$~dex) of metals in
XO-2S compared to XO-2N is plausibly due to the presence of more gas
giant planets around XO-2S, following a similar interpretation of
\citealt{Melendez:2009aa} of the trend between solar twins and the Sun.
In this scenario, forming planets in the protoplanetary disk
locks heavier elements to the core of gas giant planets.
The positive correlation of $\Delta\elemH{X} (\mathrm{N}-\mathrm{S})$
with \Tcondens\ requires a scenario involving rocky planets.
Both forming more rocky planets in XO-2S and accreting more rocky planets to
XO-2N at later stage were discussed (\citealt{Ramirez:2015aa,Biazzo:2015aa}).
The estimated mass of rocky material required to explain the observed trend
is a few tens of $\mearth$.

{\bf WASP-94A/B:}
Each star in this pair of F8 and F9 stars with super-solar metallicity
($\feh\approx 0.3$) hosts a hot Jupiter.
The planet around WASP-94A is transiting with a misaligned, probably retrograde
circular ($e<0.13$) orbit, while that hosted by WASP-94B is a little more
massive by $\sim 0.15$~\mjupiter\ and closer in, aligned with the host star.
WASP-94A shows a depletion of $0.02$~dex in volatile and moderately volatile
elements ($\Tcondens < 1200$~K) and an enhancement of $0.01$~dex in refractory
elements ($\Tcondens>1200$~K) relative to WASP-94B, with a median
uncertainty of $0.006$~dex among all elements
resulting in a statistically significant non-zero slope between
$\Delta\elemH{X}$ and $\Tcondens$ (\citealt{Teske:2016aa}).\footnote{
  Note that the condensation
  temperature $\Tcondens$ used is for solar system composition
  gas, which can differ from that of higher metallicity gas.}
Multiple possibilities related to the formation and evolution
of planetary systems around each star as well as causes unrelated to planets
such as dust cleansing during the fully convective phase or different rotation
and granulation between the stars were considered, but none was favored.

{\bf $\zeta^1/\zeta^2$ Reticuli (HD 20807/HD 20766):}
With a projected separation of $\approx 3700$~AU, both solar-type stars in this pair have
no detected planets.
However, $\zeta^2$ hosts a debris disk detected via infrared excess
(\citealt{2008ApJ...674.1086T}) as well as direct imaging
(\citealt{2010A&A...518L.131E}).
Both stars have super-solar metallicity of $\approx 0.2$~dex.
A differential abundance analysis using high-resolution spectra
shows that $\zeta^1$ is more metal rich than $\zeta^2$ by $\sim 0.02 \pm 0.003$~dex,
and that there is a positive slope between the abundance differences of 24 species
and $\Tcondens$.
A possible explanation proposed is that the relative lack of refractory elements
in $\zeta^2$ is because they are locked up in rocky bodies
that make up its debris disk (\citealt{2016A&A...588A..81S}).

{\bf HD~133131A/B:}
For this metal-poor ($\feh\approx -0.3$), old ($\sim 9.5$~Gyr) pair of
solar-type stars, high-precision radial velocity monitoring recently revealed
several planets (\citealt{Teske:2016ab}): star A hosts two eccentric giant
planets at $\approx 1.4$ and $\approx 5$~AU while star star B hosts a longer
period giant planet at $\approx 6.5$~AU.
\citealt{Teske:2016ab} measured a deficit of $0.03 \pm 0.017$~dex in
refractory elements in A relative two B without any conclusive interpretation.


\begin{thebibliography}{}
\expandafter\ifx\csname natexlab\endcsname\relax\def\natexlab#1{#1}\fi

\bibitem[{{Adams} {et~al.}(1987){Adams}, {Lada}, \&
  {Shu}}]{1987ApJ...312..788A}
{Adams}, F.~C., {Lada}, C.~J., \& {Shu}, F.~H. 1987, \apj, 312, 788

\bibitem[{{Allen} \& {Monroy-Rodr{\'{\i}}guez}(2014)}]{Allen:2014}
{Allen}, C., \& {Monroy-Rodr{\'{\i}}guez}, M.~A. 2014, \apj, 790, 158

\bibitem[{{Asplund} {et~al.}(2009){Asplund}, {Grevesse}, {Sauval}, \&
  {Scott}}]{Asplund:2009aa}
{Asplund}, M., {Grevesse}, N., {Sauval}, A.~J., \& {Scott}, P. 2009, \araa, 47,
  481

\bibitem[{{Astropy Collaboration} {et~al.}(2013){Astropy Collaboration},
  {Robitaille}, {Tollerud}, {Greenfield}, {Droettboom}, {Bray}, {Aldcroft},
  {Davis}, {Ginsburg}, {Price-Whelan}, {Kerzendorf}, {Conley}, {Crighton},
  {Barbary}, {Muna}, {Ferguson}, {Grollier}, {Parikh}, {Nair}, {Unther},
  {Deil}, {Woillez}, {Conseil}, {Kramer}, {Turner}, {Singer}, {Fox}, {Weaver},
  {Zabalza}, {Edwards}, {Azalee Bostroem}, {Burke}, {Casey}, {Crawford},
  {Dencheva}, {Ely}, {Jenness}, {Labrie}, {Lim}, {Pierfederici}, {Pontzen},
  {Ptak}, {Refsdal}, {Servillat}, \& {Streicher}}]{Astropy-Collaboration:2013}
{Astropy Collaboration}, {Robitaille}, T.~P., {Tollerud}, E.~J., {et~al.} 2013,
  \aap, 558, A33

\bibitem[{{Aumer} {et~al.}(2016){Aumer}, {Binney}, \&
  {Sch{\"o}nrich}}]{Aumer:2016}
{Aumer}, M., {Binney}, J., \& {Sch{\"o}nrich}, R. 2016, \mnras, 462, 1697

\bibitem[{{Baraffe} {et~al.}(2017){Baraffe}, {Pratt}, {Goffrey}, {Constantino},
  {Folini}, {Popov}, {Walder}, \& {Viallet}}]{Baraffe2017}
{Baraffe}, I., {Pratt}, J., {Goffrey}, T., {et~al.} 2017, \apjl, 845, L6

\bibitem[{{Battistini} \& {Bensby}(2015)}]{Battistini:2015aa}
{Battistini}, C., \& {Bensby}, T. 2015, \aap, 577, A9

\bibitem[{{Bensby} {et~al.}(2003){Bensby}, {Feltzing}, \&
  {Lundstr{\"o}m}}]{Bensby:2003aa}
{Bensby}, T., {Feltzing}, S., \& {Lundstr{\"o}m}, I. 2003, \aap, 410, 527

\bibitem[{{Biazzo} {et~al.}(2015){Biazzo}, {Gratton}, {Desidera}, {Lucatello},
  {Sozzetti}, {Bonomo}, {Damasso}, {Gandolfi}, {Affer}, {Boccato}, {Borsa},
  {Claudi}, {Cosentino}, {Covino}, {Knapic}, {Lanza}, {Maldonado}, {Marzari},
  {Micela}, {Molaro}, {Pagano}, {Pedani}, {Pillitteri}, {Piotto}, {Poretti},
  {Rainer}, {Santos}, {Scandariato}, \& {Zanmar Sanchez}}]{Biazzo:2015aa}
{Biazzo}, K., {Gratton}, R., {Desidera}, S., {et~al.} 2015, \aap, 583, A135

\bibitem[{{Bovy}(2015)}]{Bovy:2015}
{Bovy}, J. 2015, \apjs, 216, 29

\bibitem[{{Brewer} {et~al.}(2015){Brewer}, {Fischer}, {Basu}, {Valenti}, \&
  {Piskunov}}]{2015ApJ...805..126B}
{Brewer}, J.~M., {Fischer}, D.~A., {Basu}, S., {Valenti}, J.~A., \& {Piskunov},
  N. 2015, \apj, 805, 126

\bibitem[{{Brewer} {et~al.}(2016){Brewer}, {Fischer}, {Valenti}, \&
  {Piskunov}}]{2016ApJS..225...32B}
{Brewer}, J.~M., {Fischer}, D.~A., {Valenti}, J.~A., \& {Piskunov}, N. 2016,
  \apjs, 225, 32

\bibitem[{{Casey} {et~al.}(2016){Casey}, {Ruchti}, {Masseron}, {Randich},
  {Gilmore}, {Lind}, {Kennedy}, {Koposov}, {Hourihane}, {Franciosini}, {Lewis},
  {Magrini}, {Morbidelli}, {Sacco}, {Worley}, {Feltzing}, {Jeffries},
  {Vallenari}, {Bensby}, {Bragaglia}, {Flaccomio}, {Francois}, {Korn},
  {Lanzafame}, {Pancino}, {Recio-Blanco}, {Smiljanic}, {Carraro}, {Costado},
  {Damiani}, {Donati}, {Frasca}, {Jofr{\'e}}, {Lardo}, {de Laverny}, {Monaco},
  {Prisinzano}, {Sbordone}, {Sousa}, {Tautvai{\v s}ien{\.e}}, {Zaggia},
  {Zwitter}, {Delgado Mena}, {Chorniy}, {Martell}, {Silva Aguirre}, {Miglio},
  {Chiappini}, {Montalban}, {Morel}, \& {Valentini}}]{Casey:2016aa}
{Casey}, A.~R., {Ruchti}, G., {Masseron}, T., {et~al.} 2016, \mnras, 461, 3336

\bibitem[{{Chambers}(2010)}]{Chambers:2010aa}
{Chambers}, J.~E. 2010, \apj, 724, 92

\bibitem[{{Chen} {et~al.}(2001){Chen}, {Nissen}, {Benoni}, \&
  {Zhao}}]{Chen2001}
{Chen}, Y.~Q., {Nissen}, P.~E., {Benoni}, T., \& {Zhao}, G. 2001, \aap, 371,
  943

\bibitem[{{Cochran} {et~al.}(1997){Cochran}, {Hatzes}, {Butler}, \&
  {Marcy}}]{1997ApJ...483..457C}
{Cochran}, W.~D., {Hatzes}, A.~P., {Butler}, R.~P., \& {Marcy}, G.~W. 1997,
  \apj, 483, 457

\bibitem[{{Desidera} {et~al.}(2004){Desidera}, {Gratton}, {Scuderi}, {Claudi},
  {Cosentino}, {Barbieri}, {Bonanno}, {Carretta}, {Endl}, {Lucatello},
  {Martinez Fiorenzano}, \& {Marzari}}]{Desidera:2004aa}
{Desidera}, S., {Gratton}, R.~G., {Scuderi}, S., {et~al.} 2004, \aap, 420, 683

\bibitem[{{Eiroa} {et~al.}(2010){Eiroa}, {Fedele}, {Maldonado},
  {Gonz{\'a}lez-Garc{\'{\i}}a}, {Rodmann}, {Heras}, {Pilbratt}, {Augereau},
  {Mora}, {Montesinos}, {Ardila}, {Bryden}, {Liseau}, {Stapelfeldt},
  {Launhardt}, {Solano}, {Bayo}, {Absil}, {Ar{\'e}valo}, {Barrado},
  {Beichmann}, {Danchi}, {Del Burgo}, {Ertel}, {Fridlund}, {Fukagawa},
  {Guti{\'e}rrez}, {Gr{\"u}n}, {Kamp}, {Krivov}, {Lebreton}, {L{\"o}hne},
  {Lorente}, {Marshall}, {Mart{\'{\i}}nez-Arn{\'a}iz}, {Meeus}, {Montes},
  {Morbidelli}, {M{\"u}ller}, {Mutschke}, {Nakagawa}, {Olofsson}, {Ribas},
  {Roberge}, {Sanz-Forcada}, {Th{\'e}bault}, {Walker}, {White}, \&
  {Wolf}}]{2010A&A...518L.131E}
{Eiroa}, C., {Fedele}, D., {Maldonado}, J., {et~al.} 2010, \aap, 518, L131

\bibitem[{{Farihi}(2016)}]{2016NewAR..71....9F}
{Farihi}, J. 2016, \nar, 71, 9

\bibitem[{{Farihi} {et~al.}(2009){Farihi}, {Jura}, \&
  {Zuckerman}}]{2009ApJ...694..805F}
{Farihi}, J., {Jura}, M., \& {Zuckerman}, B. 2009, \apj, 694, 805

\bibitem[{{Feigelson} \& {Montmerle}(1999)}]{1999ARA&A..37..363F}
{Feigelson}, E.~D., \& {Montmerle}, T. 1999, \araa, 37, 363

\bibitem[{{Fischer} \& {Valenti}(2005)}]{Fischer:2005aa}
{Fischer}, D.~A., \& {Valenti}, J. 2005, \apj, 622, 1102

\bibitem[{Foreman-Mackey(2016)}]{corner}
Foreman-Mackey, D. 2016, The Journal of Open Source Software, 24,
  doi:10.21105/joss.00024

\bibitem[{{Foreman-Mackey} {et~al.}(2013){Foreman-Mackey}, {Hogg}, {Lang}, \&
  {Goodman}}]{2013PASP..125..306F}
{Foreman-Mackey}, D., {Hogg}, D.~W., {Lang}, D., \& {Goodman}, J. 2013, \pasp,
  125, 306

\bibitem[{{Freeman} \& {Bland-Hawthorn}(2002)}]{2002ARA&A..40..487F}
{Freeman}, K., \& {Bland-Hawthorn}, J. 2002, \araa, 40, 487

\bibitem[{{Graham} {et~al.}(1990){Graham}, {Matthews}, {Neugebauer}, \&
  {Soifer}}]{1990ApJ...357..216G}
{Graham}, J.~R., {Matthews}, K., {Neugebauer}, G., \& {Soifer}, B.~T. 1990,
  \apj, 357, 216

\bibitem[{{Gratton} {et~al.}(2001){Gratton}, {Bonanno}, {Claudi}, {Cosentino},
  {Desidera}, {Lucatello}, \& {Scuderi}}]{Gratton:2001aa}
{Gratton}, R.~G., {Bonanno}, G., {Claudi}, R.~U., {et~al.} 2001, \aap, 377, 123

\bibitem[{Hunter(2007)}]{Hunter:2007}
Hunter, J.~D. 2007, Computing In Science \& Engineering, 9, 90

\bibitem[{{Hut}(1983)}]{Hut:1983ab}
{Hut}, P. 1983, \apj, 268, 342

\bibitem[{{Hut} \& {Bahcall}(1983)}]{Hut:1983aa}
{Hut}, P., \& {Bahcall}, J.~N. 1983, \apj, 268, 319

\bibitem[{{Israelian} {et~al.}(2001){Israelian}, {Santos}, {Mayor}, \&
  {Rebolo}}]{Israelian:2001}
{Israelian}, G., {Santos}, N.~C., {Mayor}, M., \& {Rebolo}, R. 2001, \nat, 411,
  163

\bibitem[{{Jensen} {et~al.}(1996){Jensen}, {Mathieu}, \&
  {Fuller}}]{1996ApJ...458..312J}
{Jensen}, E.~L.~N., {Mathieu}, R.~D., \& {Fuller}, G.~A. 1996, \apj, 458, 312

\bibitem[{{Jiang} \& {Tremaine}(2010)}]{Jiang:2010aa}
{Jiang}, Y.-F., \& {Tremaine}, S. 2010, \mnras, 401, 977

\bibitem[{{Kilic} {et~al.}(2006){Kilic}, {von Hippel}, {Leggett}, \&
  {Winget}}]{2006ApJ...646..474K}
{Kilic}, M., {von Hippel}, T., {Leggett}, S.~K., \& {Winget}, D.~E. 2006, \apj,
  646, 474

\bibitem[{{Klein} {et~al.}(2010){Klein}, {Jura}, {Koester}, {Zuckerman}, \&
  {Melis}}]{Klein:2010aa}
{Klein}, B., {Jura}, M., {Koester}, D., {Zuckerman}, B., \& {Melis}, C. 2010,
  \apj, 709, 950

\bibitem[{{Koester} {et~al.}(2014){Koester}, {G{\"a}nsicke}, \&
  {Farihi}}]{2014A&A...566A..34K}
{Koester}, D., {G{\"a}nsicke}, B.~T., \& {Farihi}, J. 2014, \aap, 566, A34

\bibitem[{{Laws} \& {Gonzalez}(2001)}]{2001ApJ...553..405L}
{Laws}, C., \& {Gonzalez}, G. 2001, \apj, 553, 405

\bibitem[{{Liu} {et~al.}(2014){Liu}, {Asplund}, {Ram{\'{\i}}rez}, {Yong}, \&
  {Mel{\'e}ndez}}]{Liu:2014aa}
{Liu}, F., {Asplund}, M., {Ram{\'{\i}}rez}, I., {Yong}, D., \& {Mel{\'e}ndez},
  J. 2014, \mnras, 442, L51

\bibitem[{{Lodders}(2003)}]{2003ApJ...591.1220L}
{Lodders}, K. 2003, \apj, 591, 1220

\bibitem[{{Lodders} \& {Fegley}(1998)}]{1998psc..book.....L}
{Lodders}, K., \& {Fegley}, B. 1998, {The planetary scientist's companion /
  Katharina Lodders, Bruce Fegley.}

\bibitem[{{Mack} {et~al.}(2014){Mack}, {Schuler}, {Stassun}, \&
  {Norris}}]{Mack:2014aa}
{Mack}, III, C.~E., {Schuler}, S.~C., {Stassun}, K.~G., \& {Norris}, J. 2014,
  \apj, 787, 98

\bibitem[{{Mack} {et~al.}(2016){Mack}, {Stassun}, {Schuler}, {Hebb}, \&
  {Pepper}}]{Mack:2016aa}
{Mack}, III, C.~E., {Stassun}, K.~G., {Schuler}, S.~C., {Hebb}, L., \&
  {Pepper}, J.~A. 2016, \apj, 818, 54

\bibitem[{{Malmberg} {et~al.}(2011){Malmberg}, {Davies}, \&
  {Heggie}}]{Malmberg:2011aa}
{Malmberg}, D., {Davies}, M.~B., \& {Heggie}, D.~C. 2011, \mnras, 411, 859

\bibitem[{{Mason} {et~al.}(2001){Mason}, {Wycoff}, {Hartkopf}, {Douglass}, \&
  {Worley}}]{2001AJ....122.3466M}
{Mason}, B.~D., {Wycoff}, G.~L., {Hartkopf}, W.~I., {Douglass}, G.~G., \&
  {Worley}, C.~E. 2001, \aj, 122, 3466

\bibitem[{{Mayer} {et~al.}(2005){Mayer}, {Wadsley}, {Quinn}, \&
  {Stadel}}]{2005MNRAS.363..641M}
{Mayer}, L., {Wadsley}, J., {Quinn}, T., \& {Stadel}, J. 2005, \mnras, 363, 641

\bibitem[{McDonough(2001)}]{mcdonough2001composition}
McDonough, W.~F. 2001, International Geophysics, 76, 3

\bibitem[{{McDonough}(2003)}]{2003TrGeo...2..547M}
{McDonough}, W.~F. 2003, Treatise on Geochemistry, 2, 568

\bibitem[{McKinney(2010)}]{pandas}
McKinney, W. 2010, in Proceedings of the 9th Python in Science Conference, ed.
  S.~van~der Walt \& J.~Millman, 51 -- 56

\bibitem[{{Mel{\'e}ndez} {et~al.}(2009){Mel{\'e}ndez}, {Asplund}, {Gustafsson},
  \& {Yong}}]{Melendez:2009aa}
{Mel{\'e}ndez}, J., {Asplund}, M., {Gustafsson}, B., \& {Yong}, D. 2009, \apjl,
  704, L66

\bibitem[{{Melendez} \& {Ramirez}(2016)}]{2016arXiv161104064M}
{Melendez}, J., \& {Ramirez}, I. 2016, ArXiv e-prints

\bibitem[{Myles(2017 in prep)}]{jmlithium}
Myles, J. e.~a. 2017 in prep, {}

\bibitem[{{Oh} {et~al.}(2017){Oh}, {Price-Whelan}, {Hogg}, {Morton}, \&
  {Spergel}}]{2017AJ....153..257O}
{Oh}, S., {Price-Whelan}, A.~M., {Hogg}, D.~W., {Morton}, T.~D., \& {Spergel},
  D.~N. 2017, \aj, 153, 257

\bibitem[{{Paquette} {et~al.}(1986){Paquette}, {Pelletier}, {Fontaine}, \&
  {Michaud}}]{1986ApJS...61..197P}
{Paquette}, C., {Pelletier}, C., {Fontaine}, G., \& {Michaud}, G. 1986, \apjs,
  61, 197

\bibitem[{{Patience} {et~al.}(2002){Patience}, {White}, {Ghez}, {McCabe},
  {McLean}, {Larkin}, {Prato}, {Kim}, {Lloyd}, {Liu}, {Graham}, {Macintosh},
  {Gavel}, {Max}, {Bauman}, {Olivier}, {Wizinowich}, \&
  {Acton}}]{2002ApJ...581..654P}
{Patience}, J., {White}, R.~J., {Ghez}, A.~M., {et~al.} 2002, \apj, 581, 654

\bibitem[{P\'erez \& Granger(2007)}]{Perez:2007}
P\'erez, F., \& Granger, B.~E. 2007, Computing in Science and Engineering, 9,
  21

\bibitem[{{Pinsonneault}(1997)}]{1997ARA&A..35..557P}
{Pinsonneault}, M. 1997, \araa, 35, 557

\bibitem[{{Pinsonneault} {et~al.}(2001){Pinsonneault}, {DePoy}, \&
  {Coffee}}]{Pinsonneault:2001aa}
{Pinsonneault}, M.~H., {DePoy}, D.~L., \& {Coffee}, M. 2001, \apjl, 556, L59

\bibitem[{Price-Whelan {et~al.}(2017)Price-Whelan, Sipocz, \& Oh}]{gala}
Price-Whelan, A., Sipocz, B., \& Oh, S. 2017, adrn/gala: v0.1.3, , ,
  doi:10.5281/zenodo.321907

\bibitem[{{Quinn} {et~al.}(2009){Quinn}, {Wilkinson}, {Irwin}, {Marshall},
  {Koch}, \& {Belokurov}}]{Quinn:2009}
{Quinn}, D.~P., {Wilkinson}, M.~I., {Irwin}, M.~J., {et~al.} 2009, \mnras, 396,
  L11

\bibitem[{{Ram{\'{\i}}rez} {et~al.}(2012){Ram{\'{\i}}rez}, {Fish}, {Lambert},
  \& {Allende Prieto}}]{2012ApJ...756...46R}
{Ram{\'{\i}}rez}, I., {Fish}, J.~R., {Lambert}, D.~L., \& {Allende Prieto}, C.
  2012, \apj, 756, 46

\bibitem[{{Ram{\'{\i}}rez} {et~al.}(2011){Ram{\'{\i}}rez}, {Mel{\'e}ndez},
  {Cornejo}, {Roederer}, \& {Fish}}]{2011ApJ...740...76R}
{Ram{\'{\i}}rez}, I., {Mel{\'e}ndez}, J., {Cornejo}, D., {Roederer}, I.~U., \&
  {Fish}, J.~R. 2011, \apj, 740, 76

\bibitem[{{Ram{\'{\i}}rez} {et~al.}(2015){Ram{\'{\i}}rez}, {Khanal}, {Aleo},
  {Sobotka}, {Liu}, {Casagrande}, {Mel{\'e}ndez}, {Yong}, {Lambert}, \&
  {Asplund}}]{Ramirez:2015aa}
{Ram{\'{\i}}rez}, I., {Khanal}, S., {Aleo}, P., {et~al.} 2015, \apj, 808, 13

\bibitem[{{Rasio} \& {Ford}(1996)}]{1996Sci...274..954R}
{Rasio}, F.~A., \& {Ford}, E.~B. 1996, Science, 274, 954

\bibitem[{{Reach} {et~al.}(2005){Reach}, {Kuchner}, {von Hippel}, {Burrows},
  {Mullally}, {Kilic}, \& {Winget}}]{2005ApJ...635L.161R}
{Reach}, W.~T., {Kuchner}, M.~J., {von Hippel}, T., {et~al.} 2005, \apjl, 635,
  L161

\bibitem[{{Reddy} {et~al.}(2002){Reddy}, {Lambert}, {Laws}, {Gonzalez}, \&
  {Covey}}]{2002MNRAS.335.1005R}
{Reddy}, B.~E., {Lambert}, D.~L., {Laws}, C., {Gonzalez}, G., \& {Covey}, K.
  2002, \mnras, 335, 1005

\bibitem[{{Saffe} {et~al.}(2015){Saffe}, {Flores}, \& {Buccino}}]{Saffe:2015aa}
{Saffe}, C., {Flores}, M., \& {Buccino}, A. 2015, \aap, 582, A17

\bibitem[{{Saffe} {et~al.}(2016){Saffe}, {Flores}, {Jaque Arancibia},
  {Buccino}, \& {Jofr{\'e}}}]{2016A&A...588A..81S}
{Saffe}, C., {Flores}, M., {Jaque Arancibia}, M., {Buccino}, A., \&
  {Jofr{\'e}}, E. 2016, \aap, 588, A81

\bibitem[{{Santos} {et~al.}(2004){Santos}, {Israelian}, \&
  {Mayor}}]{Santos2004}
{Santos}, N.~C., {Israelian}, G., \& {Mayor}, M. 2004, \aap, 415, 1153

\bibitem[{{Sch{\"o}nrich}(2012)}]{Schonrich:2012}
{Sch{\"o}nrich}, R. 2012, \mnras, 427, 274

\bibitem[{{Sch{\"o}nrich} {et~al.}(2010){Sch{\"o}nrich}, {Binney}, \&
  {Dehnen}}]{Schonrich:2010}
{Sch{\"o}nrich}, R., {Binney}, J., \& {Dehnen}, W. 2010, \mnras, 403, 1829

\bibitem[{{Schuler} {et~al.}(2011){Schuler}, {Cunha}, {Smith}, {Ghezzi},
  {King}, {Deliyannis}, \& {Boesgaard}}]{2011ApJ...737L..32S}
{Schuler}, S.~C., {Cunha}, K., {Smith}, V.~V., {et~al.} 2011, \apjl, 737, L32

\bibitem[{{Spada} {et~al.}(2013){Spada}, {Demarque}, {Kim}, \&
  {Sills}}]{2013ApJ...776...87S}
{Spada}, F., {Demarque}, P., {Kim}, Y.-C., \& {Sills}, A. 2013, \apj, 776, 87

\bibitem[{{Teske} {et~al.}(2015){Teske}, {Ghezzi}, {Cunha}, {Smith}, {Schuler},
  \& {Bergemann}}]{Teske:2015aa}
{Teske}, J.~K., {Ghezzi}, L., {Cunha}, K., {et~al.} 2015, \apjl, 801, L10

\bibitem[{{Teske} {et~al.}(2016{\natexlab{a}}){Teske}, {Khanal}, \&
  {Ram{\'{\i}}rez}}]{Teske:2016aa}
{Teske}, J.~K., {Khanal}, S., \& {Ram{\'{\i}}rez}, I. 2016{\natexlab{a}}, \apj,
  819, 19

\bibitem[{{Teske} {et~al.}(2013){Teske}, {Schuler}, {Cunha}, {Smith}, \&
  {Griffith}}]{Teske:2013aa}
{Teske}, J.~K., {Schuler}, S.~C., {Cunha}, K., {Smith}, V.~V., \& {Griffith},
  C.~A. 2013, \apjl, 768, L12

\bibitem[{{Teske} {et~al.}(2016{\natexlab{b}}){Teske}, {Shectman}, {Vogt},
  {D{\'{\i}}az}, {Butler}, {Crane}, {Thompson}, \& {Arriagada}}]{Teske:2016ab}
{Teske}, J.~K., {Shectman}, S.~A., {Vogt}, S.~S., {et~al.} 2016{\natexlab{b}},
  \aj, 152, 167

\bibitem[{{Th{\'e}venin} {et~al.}(2017){Th{\'e}venin}, {Oreshina}, {Baturin},
  {Gorshkov}, {Morel}, \& {Provost}}]{Thevenin2017}
{Th{\'e}venin}, F., {Oreshina}, A.~V., {Baturin}, V.~A., {et~al.} 2017, \aap,
  598, A64

\bibitem[{{Trilling} {et~al.}(2008){Trilling}, {Bryden}, {Beichman}, {Rieke},
  {Su}, {Stansberry}, {Blaylock}, {Stapelfeldt}, {Beeman}, \&
  {Haller}}]{2008ApJ...674.1086T}
{Trilling}, D.~E., {Bryden}, G., {Beichman}, C.~A., {et~al.} 2008, \apj, 674,
  1086

\bibitem[{{Tucci Maia} {et~al.}(2014){Tucci Maia}, {Mel{\'e}ndez}, \&
  {Ram{\'{\i}}rez}}]{2014ApJ...790L..25T}
{Tucci Maia}, M., {Mel{\'e}ndez}, J., \& {Ram{\'{\i}}rez}, I. 2014, \apjl, 790,
  L25

\bibitem[{{Udry} {et~al.}(2017){Udry}, {Dumusque}, {Lovis}, {Segransan},
  {Diaz}, {Benz}, {Bouchy}, {Coffinet}, {Lo Curto}, {Mayor}, {Mordasini},
  {Motalebi}, {Pepe}, {Queloz}, {Santos}, {Wyttenbach}, {Alonso}, {Collier
  Cameron}, {Deleuil}, {Figueira}, {Gillon}, {Moutou}, {Pollacco}, \&
  {Pompei}}]{2017arXiv170505153U}
{Udry}, S., {Dumusque}, X., {Lovis}, C., {et~al.} 2017, ArXiv e-prints

\bibitem[{{Van der Walt} {et~al.}(2011){Van der Walt}, Colbert, \&
  Varoquaux}]{Van-der-Walt:2011}
{Van der Walt}, S., Colbert, S.~C., \& Varoquaux, G. 2011, {Computing in
  Science \& Engineering}, 13, 22

\bibitem[{{VandenBerg} \& {Clem}(2003)}]{2003AJ....126..778V}
{VandenBerg}, D.~A., \& {Clem}, J.~L. 2003, \aj, 126, 778

\bibitem[{{Vanderburg} {et~al.}(2015){Vanderburg}, {Johnson}, {Rappaport},
  {Bieryla}, {Irwin}, {Lewis}, {Kipping}, {Brown}, {Dufour}, {Ciardi}, {Angus},
  {Schaefer}, {Latham}, {Charbonneau}, {Beichman}, {Eastman}, {McCrady},
  {Wittenmyer}, \& {Wright}}]{2015Natur.526..546V}
{Vanderburg}, A., {Johnson}, J.~A., {Rappaport}, S., {et~al.} 2015, \nat, 526,
  546

\bibitem[{{Weidenschilling} \& {Marzari}(1996)}]{1996Natur.384..619W}
{Weidenschilling}, S.~J., \& {Marzari}, F. 1996, \nat, 384, 619

\bibitem[{{Wielen}(1977)}]{Wielen:1977}
{Wielen}, R. 1977, \aap, 60, 263

\bibitem[{{Yoo} {et~al.}(2004){Yoo}, {Chanam{\'e}}, \& {Gould}}]{Yoo:2004aa}
{Yoo}, J., {Chanam{\'e}}, J., \& {Gould}, A. 2004, \apj, 601, 311

\bibitem[{{Zakamska} \& {Tremaine}(2004)}]{Zakamska:2004aa}
{Zakamska}, N.~L., \& {Tremaine}, S. 2004, \aj, 128, 869

\bibitem[{{Zuckerman} \& {Becklin}(1987)}]{1987Natur.330..138Z}
{Zuckerman}, B., \& {Becklin}, E.~E. 1987, \nat, 330, 138

\bibitem[{{Zuckerman} {et~al.}(2007){Zuckerman}, {Koester}, {Melis}, {Hansen},
  \& {Jura}}]{Zuckerman:2007aa}
{Zuckerman}, B., {Koester}, D., {Melis}, C., {Hansen}, B.~M., \& {Jura}, M.
  2007, \apj, 671, 872

\bibitem[{{Zuckerman} {et~al.}(2003){Zuckerman}, {Koester}, {Reid}, \&
  {H{\"u}nsch}}]{2003ApJ...596..477Z}
{Zuckerman}, B., {Koester}, D., {Reid}, I.~N., \& {H{\"u}nsch}, M. 2003, \apj,
  596, 477

\bibitem[{{Zuckerman} {et~al.}(2010){Zuckerman}, {Melis}, {Klein}, {Koester},
  \& {Jura}}]{2010ApJ...722..725Z}
{Zuckerman}, B., {Melis}, C., {Klein}, B., {Koester}, D., \& {Jura}, M. 2010,
  \apj, 722, 725

\end{thebibliography}
\end{document}